\documentclass[twocolumn, times, tighten]{aastex62}

\newcommand{\reply}{}
\newcommand{\resp}{}

\accepted{23 April 2018}
\submitjournal{ApJ}
\shorttitle{EX~Lup Gas Disk}
\shortauthors{Hales et al.}

\begin{document}
\title{The Circumstellar Disk and Asymmetric outflow of the EX Lup Outburst System}

\author{A.S. Hales}
\affiliation{Joint ALMA Observatory, Avenida Alonso de C\'ordova 3107, Vitacura 7630355, Santiago, Chile}
\affiliation{National Radio Astronomy Observatory, 520 Edgemont Road, Charlottesville, VA 22903-2475, United States of America}
\author{S. P\'erez}
\affiliation{Departamento de Astronom\'ia, Universidad de Chile, Casilla 36-D, Santiago 8330015, Chile}
\author{M. Saito}
\affiliation{National Astronomical Observatory of Japan, Osawa 2-21-1, Mitaka, Tokyo 181-8588, Japan}
\affiliation{Department of Astronomy, The Graduate University for Advanced Studies (SOKENDAI), Osawa 2-21-1, Mitaka, Tokyo 181-8588, Japan}
\author{C. Pinte}
\affiliation{Universit{\'e} Grenoble Alpes, CNRS, IPAG, F-38000 Grenoble, France}
\author{L.B.G. Knee}
\affiliation{NRC Herzberg Astronomy and Astrophysics Research Centre, 5071 West Saanich Road, Victoria, BC V9E 2E7, Canada}
\author{I. de Gregorio-Monsalvo}
\affiliation{Joint ALMA Observatory, Avenida Alonso de C\'ordova 3107, Vitacura 7630355, Santiago, Chile}
\affiliation{European Southern Observatory, Avenida Alonso de C\'{o}rdova 3107, Vitacura 7630355, Santiago, Chile}
\author{B. Dent}
\affiliation{Joint ALMA Observatory, Avenida Alonso de C\'ordova 3107, Vitacura 7630355, Santiago, Chile}
\affiliation{European Southern Observatory, Avenida Alonso de C\'{o}rdova 3107, Vitacura 7630355, Santiago, Chile}
\author{C. L\'opez}
\affiliation{Joint ALMA Observatory, Avenida Alonso de C\'ordova 3107, Vitacura 7630355, Santiago, Chile}
\author{A. Plunkett}
\affiliation{European Southern Observatory, Avenida Alonso de C\'{o}rdova 3107, Vitacura 7630355, Santiago, Chile}
\author{P. Cort\'es}
\affiliation{Joint ALMA Observatory, Avenida Alonso de C\'ordova 3107, Vitacura 7630355, Santiago, Chile}
\affiliation{National Radio Astronomy Observatory, 520 Edgemont Road, Charlottesville, VA 22903-2475, United States of America}
\author{S. Corder}
\affiliation{Joint ALMA Observatory, Avenida Alonso de C\'ordova 3107, Vitacura 7630355, Santiago, Chile}
\affiliation{National Radio Astronomy Observatory, 520 Edgemont Road, Charlottesville, VA 22903-2475, United States of America}
\author{L. Cieza}
\affiliation{Nucleo de Astronom\'ia, Facultad de Ingenier\'ia y Ciencias, Universidad Diego Portales, Av. Ejercito 441, Santiago, Chile}
\affiliation{Millenium Nucleus Protoplanetary Discs in ALMA Early Science, Universidad Diego Portales, Av. Ejercito 441, Santiago, Chile}


\correspondingauthor{A.S. Hales}
\email{antonio.hales@alma.cl}


\begin{abstract}


  We present Atacama Large Millimeter/submillimeter Array (ALMA)
  observations at 0.3$\arcsec$-resolution of EX Lup, the prototype of
  the EXor class of outbursting pre-main sequence stars. The
  circumstellar disk of EX Lup is resolved for the first time in 1.3
  mm continuum emission and in the $J$=2--1 spectral line of three
  isotopologues of CO. At the spatial resolution and sensitivity
  achieved, the compact dust continuum disk shows no indications of
  clumps, fragments, or asymmetries \reply{above 5-sigma
    level}. Radiative transfer modeling constrains the characteristic
  radius of the dust disk to 23\,au and a total dust mass of
  1.0$\times$10$^{-4}$\,M$_\odot$ (33\,M$_\earth$), similar to other
  EXor sources. The $^{13}$CO and C$^{18}$O line emission trace the
  disk rotation and are used to constrain the disk geometry,
  kinematics, and a total gas disk mass of
  \reply{5.1$\times$10$^{-4}$\,M$_\odot$}. The $^{12}$CO emission
  extends out to a radius of 200\,au and is asymmetric, with one side
  deviating from Keplerian rotation. We detect blue-shifted, $^{12}$CO
  arc-like emission located 0.8\arcsec to the north-west, and
  spatially disconnected from the disk emission. We interpret this
  extended structure as the brightened walls of a cavity excavated by
  an outflow, which are more commonly seen in FUor sources. Such
  outflows have also been seen in the borderline FU/EXor object V1647
  Ori, but not towards EXor objects. Our detection provides evidence
  that the outflow phenomenon persists into the EXor phase, suggesting
  that FUor and EXor objects are a continuous population in which
  outflow activity declines with age, with transitional objects such
  as EX Lup and V1647 Ori.
  


\end{abstract}



\section{Introduction}\label{sec:intro}

Most young stars are now thought to undergo phases of episodic
accretion during which their accretion rates can increase by several
orders of magnitude \citep{hartmann1996}. Although short-lived
compared to the total life of the accretion disks, these bursts are
believed to play a key role in the building up of the final stellar
mass \citep{hartmann2008,hartmann2016}. These episodic accretion
events may alleviate the well-known luminosity problem, in which
low-mass stars appear under-luminous compared to the predictions of
steady-state accretion models \citep{kenyon1990,evans2009}. The
enhancement in accretion luminosity can affect the dust and gas
properties and the variable snow-lines can influence the disk
chemistry, dust grain properties, re-condensation of ices, and
possibly planet formation
\citep{2009Natur.459..224A,banzatti2015,cieza2016,hubbard2017,rab2017}.

Outburst sources have been historically divided into two classes,
FUors (named after their prototype FU~Ori) and EXors (named after
their prototype EX~Lup). The class distinction is observational: FUors
have large ($>$~5~mag), long-lived (years to decades) bursts
\citep{herbig1966}, whereas EXors have moderate (2--4~mag), shorter
(days to months) episodes of high accretion. \resp{Recent observations
  have questioned the need for two separate classification, leading to
  the suggestion that both types of outburst may have a common origin
  \citep[][and references therein]{audard2014}.}

Different possible triggering mechanisms have
been proposed for the episodic outburst: (1) thermal instability
\citep{bell1994}, (2) coupling of the gravitational (GI) and
magnetorotational (MRI) instability \citep{armitage2001,zhu2009}, (3)
disk fragmentation \citep{vorobyov2015}, and (4) tidal interactions
between the disk and either planets \citep{lodato2004} or stellar
companions \citep{Bonnell1992}. \citet{dangelo2010} showed that disk
material near corotation radius can become trapped, which could also
produce variable accretion rates. Recently, \citet{armitage2016}
proposed that EXor outbursts could be produced by the interaction
between stellar dynamo cycles and the inner regions of the disk.

Some proposed mechanisms for FUor and EXor outbursts (e.g., GI and
disk fragmentation) predict the presence of distinct morphological
features that might be detectable in the dust continuum of massive
disks, yet to date none of these signatures have been found in high
resolution interferometric images of FUor/EXor disks
\citep{hales2015,cieza2016,ruiz2017,zurlo2017,kospal2017b,principe2018}. Even
in V883~Ori, the most massive FUor disk known and thus the best
candidate for a GI to occur, ALMA observations at 0.03$\arcsec$
(12~au) resolution were unable to detect the predicted signatures of
disk instability or fragmentation \citep{cieza2016}. Observations of
FUor/EXor sources show that the continuum disks are compact, with
characteristic radii ${\rm R}_{\rm c}~<~$20--40~au \citep{cieza2017}
and most likely optically thick in their inner regions -- making the
total dust mass difficult to estimate \citep{cieza2016,liu2017}.

Complementary to dust continuum observations are spectroscopic studies of the gas
content of the disk, which provides information on both the structure and kinematics
of the system. For
example, deviations from Keplerian motion may be indicative of disk instabilities or
enhanced accretion \citep{ilee2011,evans2015}. Several physical mechanisms can produce
non-Keplerian radial velocity profiles, such as 1) radial inflows, 2) a physical warp
in the inner disk, 3) circumbinary disk interactions, and 4) a
perturbation from a stellar or planetary companion
\citep{casassus2015,rosenfeld2014,takakuwa2017}. All these processes
can be linked to enhanced stellar accretion, therefore detecting such signatures in
spectral line images can provide important clues to the physical
processes which drive episodic accretion in young stars. Furthermore,
observation of mass inflow onto the star may help constrain estimates of the
quantity of mass being transferred during episodic accretion events.

Spectral line observations show that all studied FUor sources have
active circumstellar environments characterized by strong outflow activity
\citep{ruiz2017,zurlo2017,kospal2017b}. On the other hand, EXor
sources do not show detectable outflows, with the possible exception of
V1647~Ori -- a system with an ambiguous FUor/EXor
classification \citep{principe2018}. There is also observational
evidence that most FUor sources are surrounded by large envelopes
that are still transferring material onto the disk
\citep[e.g.][and references therein]{kospal2017a,feher2017}. It has been proposed
that the imbalance between mass transferred from the envelope to the
disk and the accretion of disk mass into the star may trigger the
disk instability responsible for the outburst
\citep{bell1994,goto2011,kospal2017b}. Studying the large scale
structure (outer disk, envelope, outflows) as well as the
sub-structures (disk inner regions, clumps) of eruptive sources is
therefore crucial for understanding the nature of the FUor/EXor outbursts.

\subsection*{EX Lup}

EX~Lup is a young M0 star located in the Lupus~3 cloud
\citep[d = 155~pc;][]{gras2005,lombardi2008,comeron2008} and is the
prototype of the EXor class of young eruptive stars
\citep{herbig1950,herbig1989,herbig2008}. EX~Lup has undergone several
outbursts, with the two strongest observed in 1955 \citep{herbig1977}
and 2008 \citep{jones2008}. These outbursts are usually
ascribed to episodes of mass infall up to three orders of magnitude higher than
typical quiescent accretion rates \citep{audard2014}. The
inner disk of EX~Lup has been studied thoroughly using spectroscopic
techniques, which has shown strong variations in the physical conditions of the
gaseous and silicate components in the inner $<$~0.4~au of the disk during and
post-outburst \citep{goto2011,kospal2011,juhasz2012,Sicilia2012,Sicilia2015}. Studies
of ionized metal lines have been used to probe accretion processes in EX~Lup. These
line profiles appear to arise in a hot, asymmetric accreting region located in the
inner disk regions 0.1--0.2~au from the star \citep{Sicilia2012}, and provides
evidence for a strong inner disk wind that increases during the periods of high
accretion. Observations of the vibrationally excited mid-infrared CO lines
also indicate a strong inner disk asymmetry, attributed to the
presence of a disk hot spot, as well as the presence of a CO disk wind
\citep{goto2011}.

However, very little is known about the outer disk of EX~Lup since it
has never been spatially resolved at any wavelength. There is evidence
for a 0.025~M$_{\odot}$ and 150~au disk from spectral energy
distribution (SED) modeling \citep{sipos2009}. The SED of EX~Lup
resembles that of a Class~II object, suggesting that the circumstellar
environment consists of a disk with possibly a thin envelope. In that
sense, despite its outburst activity \reply{and young age
\citep[0.5~Myr;][]{frasca2017}}, EX~Lup appears to be relatively
evolved among eruptive young stellar objects. Recently,
\citet{kospal2016b} detected molecular emission from EX~Lup in the
$^{12}$CO(3--2), $^{12}$CO(4--3) and $^{13}$CO(3--2) lines. They
detected single-peaked line profiles which they modeled as a Keplerian
disk. The total disk gas mass inferred from the $^{13}$CO data
(2.3$\times$10$^{-4}$~M$_\odot$) is significantly lower than the one
derived from the continuum data (assuming a canonical gas-to-dust
ratio of 100), indicative of CO depletion. Despite these efforts, the
basic disk parameters of the dust and gas disk in EX~Lup remain poorly
constrained in the absence of spatially resolved images.
 
In this work we present ALMA observations that resolve the disk around
EX~Lup in both dust and gas. Section~\ref{obs} describes the details
of the ALMA observations and data reduction. In Section~\ref{results}
we present the results and an analysis of the data. A comparison of
these results with radiative transfer modeling is discussed in
Section~\ref{model} with a full discussion of our results in
Section~\ref{discussion}.  Our conclusions are summarized in
Section~\ref{conclusion}.

\section{Observations}\label{obs}

ALMA observations of EX~Lup using the facility Band~6 ($\sim$~230~GHz) receivers were
acquired on 25~July 2016 (during ALMA Cycle~3) utilizing interferometric observations
on 40 antennas in the main 12-meter antenna array, providing baselines ranging from
15.1~m to 1.12~km. For Band 6 this antenna configuration yields an angular resolution
of $\sim$~0.3$\arcsec$ ($\sim$~45~au) and a Maximum Recoverable Scale (MRS; see
\citealt{schieven2017}) of $\sim$~3.2$\arcsec$ ($\sim$~500~au).
During the observations the
precipitable water vapor column in the atmosphere was stable at 0.8--0.9~mm with clear
sky conditions, resulting in median system temperature of 70~K.

Four spectral windows were positioned to target the $^{12}$CO(2--1),
$^{13}$CO(2--1), C$^{18}$O(2--1) and the vibrationally excited
$^{13}$CO(2--1, v = 1) transitions of carbon monoxide (rest
frequencies of 230.538~GHz, 220.399~GHz, 219.560~GHz, and 218.437~GHz,
respectively). The ALMA correlator was configured in Frequency
Division Mode (FDM) to provide spectral resolutions of 61.035~kHz for
$^{12}$CO(2--1) and 122.070~kHz for the other transitions,
corresponding to velocity resolutions of 0.092 and 0.192~km~s$^{-1}$
respectively. A fifth spectral window configured in Time Division Mode
(TDM) was positioned in a region devoid of line emission for detecting
the continuum dust emission (centered at 232.477~GHz/1.29~mm). The
effective bandwidth for each spectral window was 117.187~MHz for
$^{12}$CO(2--1), $^{13}$CO(2--1), and C$^{18}$O(2--1), 234.375~MHz for
$^{13}$CO(2--1, v = 1), and \reply{1.875~GHz for the continuum}.

The primary flux calibrator was the quasar J1427$-$4206, which was also used as the
bandpass calibrator. The flux density of the amplitude
calibrator was obtained as a result of interpolation between Band~3
and Band~7 monitoring observations, both acquired two days before the
target observation (the monitor observations used Callisto as the main flux
calibrator). The Band~3 to Band~7 spectral index used for the interpolation was
-0.54 (uncertainty $\pm~$0.05). The quasar J1610$-$3958 was observed for phase
calibration. Observations of the phase calibrator, located
1.4$^{\circ}$ from the target, were alternated with the science
target every 6 minutes to calibrate the time-dependent variation of
the complex gains. The total time spent on-source was 28.3 minutes. A
secondary phase calibrator, J1604$-$4228, was observed regularly as
a check source. The purpose of the check source was to assess the
quality of the phase transfer (i.e. to estimate the coherence of the
calibrated data during the Quality Assurance data processing).

All data were calibrated using the ALMA Science Pipeline (version
r36660 of Pipeline-Cycle3-R4-B release) in the
\citep[CASA{\footnote{\url{http://casa.nrao.edu/}}};][]{2007ASPC..376..127M}
package by the East Asia ALMA ARC staff. The Pipeline uses CASA tasks
wherever possible to perform the data reduction and calibration in a
standard fashion, which included offline Water Vapor Radiometer (WVR)
calibration, system temperature correction, as well as bandpass,
phase, and amplitude calibrations. Online flagging and nominal
flagging, such as for shadowed antennas and band edges, were applied
for calibration. In addition, some scans of the phase calibrator for
six antennas (DA55, DA62, DV16, DV19, DV21, DV25) and the adjacent
target scans were flagged due to large phase scatter and/or
discontinuous phase of the calibrator.

Imaging of the continuum and molecular emission lines was performed
using the {\sc clean} task in CASA. The 1.875-GHz wide, line-free,
spectral window centered at 232.477~GHz was imaged using Briggs
weighting with a robust parameter of 0.5, resulting in a synthesized
beam size of 0.32$\arcsec\times$0.26$\arcsec$ at position angle PA =
61.8$^{\circ}$ in the final continuum image. A single iteration of
phase-only self-calibration was performed to improve the coherence in
the final image. \reply{The signal-to-noise ratio (SNR) improved by a
  factor of 1.8 after applying self-calibration.} A second continuum
image was produced at 218.474~GHz (1.37~mm) using the 234.375~MHz-wide
$^{13}$CO(2--1, v = 1) spectral window using similar {\sc clean}
parameters as for the 1.29~mm image.

Continuum subtraction in the visibility domain was performed prior to
imaging of each molecular line. Self-calibration tables derived from
the continuum were applied to the FDM spectral windows before imaging
the CO lines. During {\sc clean}ing, the spectral axis was binned to
0.2~km~s$^{-1}$ per channel for $^{12}$CO(2--1), to 0.3~km~s$^{-1}$
for $^{13}$CO(2--1), and to 0.8~km~s$^{-1}$ for C$^{18}$O(2--1) and
$^{13}$CO(2--1, v = 1), to increase the SNR. {\sc clean} was run with
natural weighting to maximize sensitivity. A root-mean-squared (RMS)
noise level of 4.5~mJy\,beam$^{-1}$ per 0.2~km~s$^{-1}$ channel was
reached in the $^{12}$CO(2--1) line, with a
0.39$\arcsec\times$0.29$\arcsec$ synthesized beam (PA =
62.7$^{\circ}$). The RMS noise per channel in the $^{13}$CO(2--1) line
was found to range between 3 and 4~mJy\,beam$^{-1}$ \reply{, with a
  mean value of 3.6~mJy\,beam$^{-1}$}.  For the C$^{18}$O(2--1) and
$^{13}$CO(2--1, v = 1) cubes, the RMS ranged between 2 and
3~mJy\,beam$^{-1}$ \reply{, with mean values of 2.2~mJy\,beam$^{-1}$
  and 2.0~mJy\,beam$^{-1}$} respectively. A second $^{12}$CO(2--1)
line image was created using Briggs weighting with a robust parameter
of 0.5, which provided a resolution of
0.32$\arcsec\times$0.26$\arcsec$ (PA = 61.8$^{\circ}$).

\begin{figure*}
\centering\includegraphics[width=0.9\textwidth]{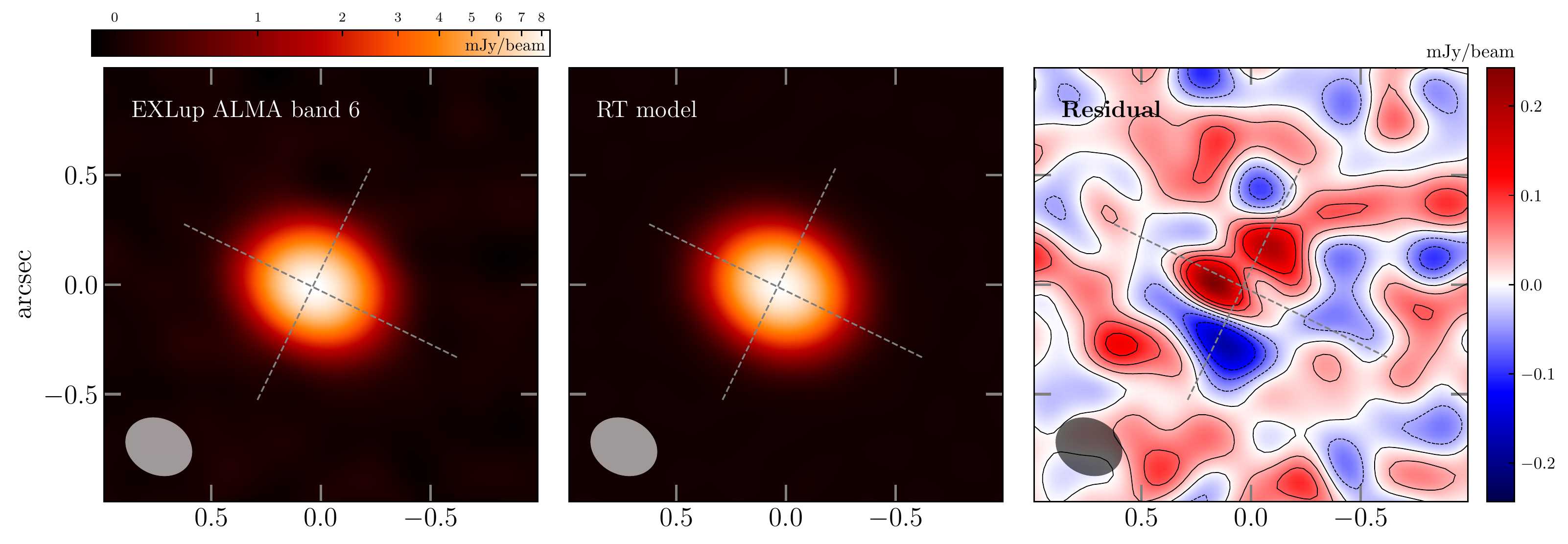}
\caption{{\it{Left-panel:}} ALMA~1.29~mm continuum image of the EX~Lup
  disk. {\it{Center-panel:}} {\sc clean} image of the best-fit model
  to the EX~Lup continuum data (as described in Section~\ref{models})
  after filtering through the observed visibilities using CASA task
  {\sc ft} (version~5.0). {\it{Right-panel:}} Residual image obtained
  by subtracting the model from the data.  \reply{The color maps in
    the left and middle panels are normalized to an asinh scale. The
    residual image on the right has maximum and minimum values equal
    to 0.24 and -0.18~mJy$\,$beam$^{-1}$, respectively. Contours show
    residual levels at -3, -2, -1, 0, 1, 2, 3 times the RMS of
    38~$\mu$Jy$\,$beam$^{-1}$. The residuals suggest the presence of
    structure not accounted for by our axisymmetric model, at the
    5~sigma level.}  }\label{fig-0}
\end{figure*}

\section{Results and analysis}\label{results}
\subsection{Dust Continuum Emission}

The disk around EX~Lup is detected and resolved in the dust
continuum. Fig.~\ref{fig-0} shows the 1.29~mm compact source, which
is detected at a SNR of 226 (peak flux 8.8~mJy\,beam$^{-1}$ with noise RMS
0.038~mJy\,beam$^{-1}$. The peak of the continuum
emission coincides with the near-infrared position of EX~Lup
\citep[as listed in the {\it{2MASS}} catalogue, after correcting for the proper
motion;][]{2003yCat.2246....0C}. The parameters of the compact
continuum emission can be determined from elliptical Gaussian fitting in the
image domain using the CASA task {\sc imfit}. The parameters of the
elliptical Gaussian component determined using {\sc imfit} are detailed in
Table~\ref{table-1}. A 0.45$\arcsec\times$0.37$\arcsec$ (FWHM) disk at
PA = 63.1$^{\circ}~\pm$~1.2$^{\circ}$ is resolved, with a total 1.29~mm
flux of 17.37$~\pm~$0.15~mJy (uncertainties are derived
from the {\sc imfit} task). An estimate of the deconvolved disk size is
0.31$\arcsec\times$0.26$\arcsec$ at PA = 64.6$^{\circ}~\pm~3.3^{\circ}$.
There is no evidence of residual emission \reply{above
5-$\sigma$} after subtracting the best-fit elliptical Gaussian from the continuum
image.

Assuming that the deprojected EX~Lup disk is circular, the major/minor
axis ratio implies a disk inclination angle of
32.4$^{\circ}~\pm~$0.9$^{\circ}$. \reply{This inclination is in
  reasonable agreement with the value of 40-50$^{\circ}$ derived by
  \citet{goto2011}, and with the 40$^{\circ}$ inclination determined
  by \citet{kospal2014}.}  It should be noted that CASA documentation
indicates that current limitations in {\sc imfit} generally results in
it underestimating the uncertainty in output axis size and PA fit
values.



The total flux measured in the 234.375~MHz-wide continuum window at
1.37~mm is 15.16$~\pm~$0.32~mJy (the RMS noise was
0.1~mJy\,beam$^{-1}$). This results in a spectral index
(F$_\nu\propto\nu^\alpha$) between 1.30 and 1.37~mm of
2.19$~\pm~$0.47. Since the Band~3 and Band~7 observations of the flux
calibrator were obtained simultaneously, and only two days apart from
the observations of EX~Lup, the effect of uncertainties in the
spectral slope of the calibrator on the in-band spectral index of
EX~Lup are minimal. The derived spectral index is in agreement with
previous estimates of the 1--3~mm slope of 2.6$~\pm~$0.4 made with
ATCA \citep{ubach2012}. Assuming that all the flux detected at 0.87~mm
with APEX/LABOCA comes from the disk
\citep[41$~\pm~$10~mJy;][]{juhasz2012}, the derived 0.87/1.3~mm
spectral index is 2.18, almost identical to the 1.3~mm spectral index
we measure in-band.


%
%
%
%
%
%
%
%
%
%
%
%
%

\subsection{Spectral Line Emission}

\begin{figure*}
\includegraphics[width=0.33\textwidth]{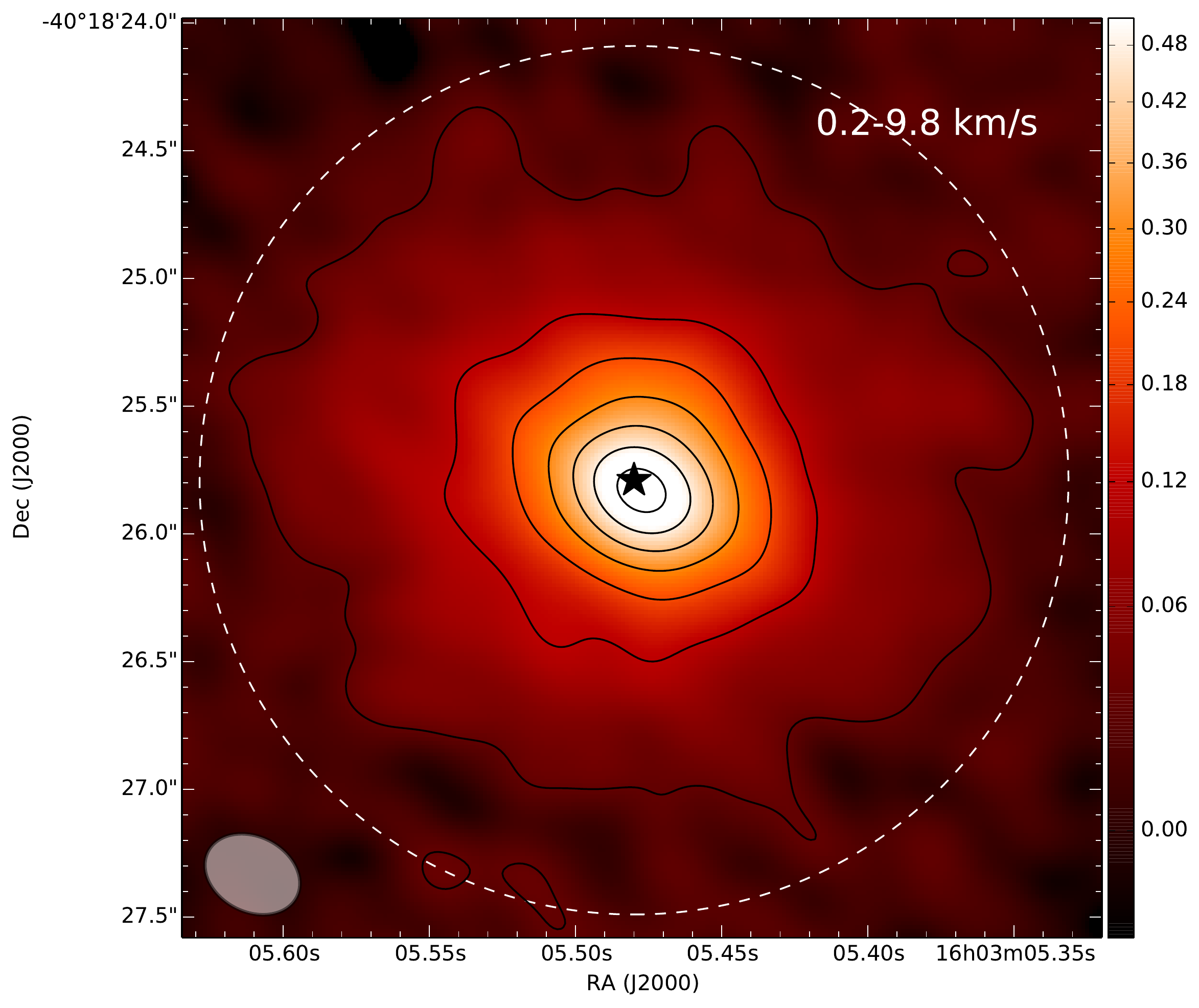}
\includegraphics[width=0.33\textwidth]{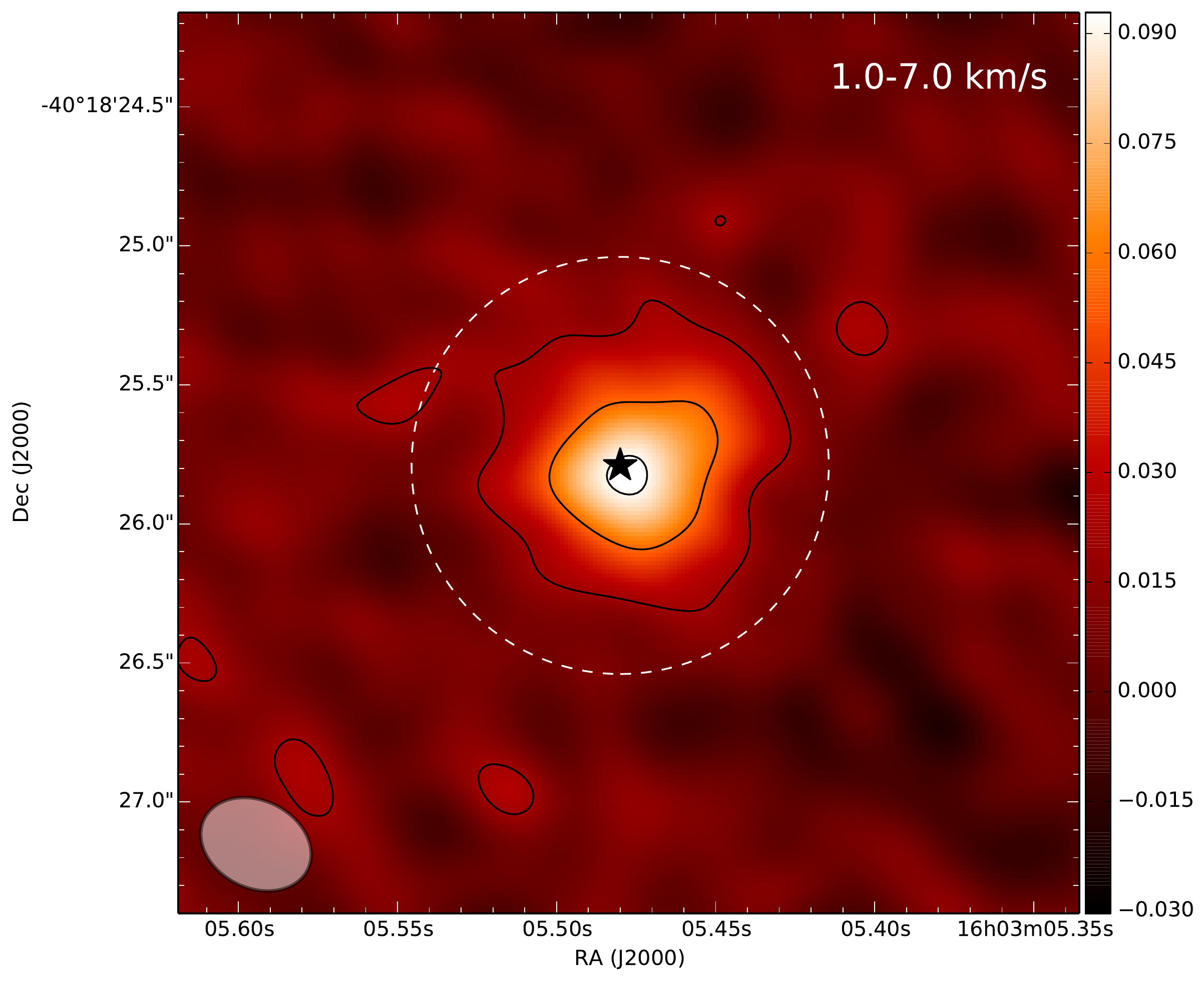}
\includegraphics[width=0.33\textwidth]{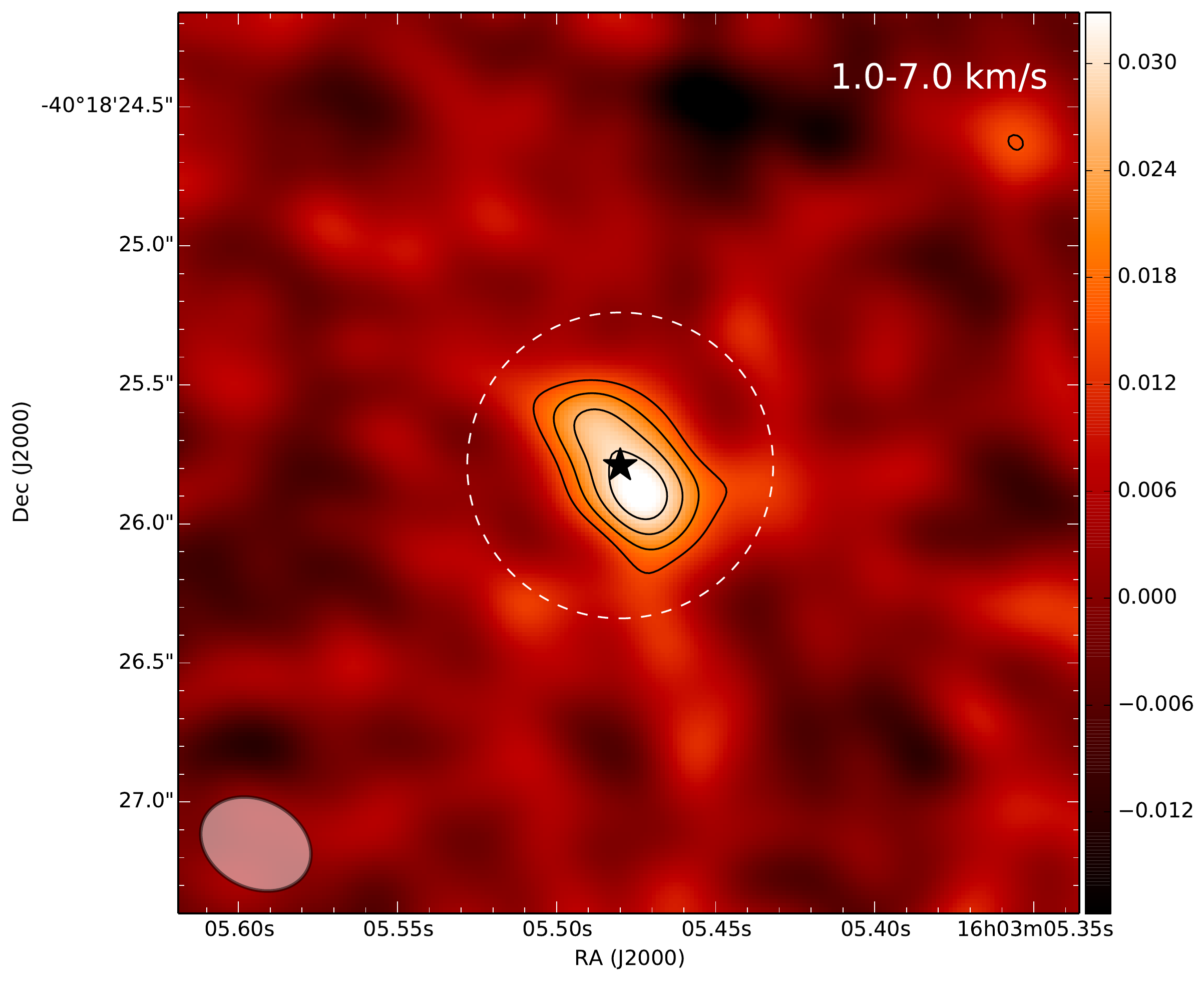}
\caption{
{\it Left-panel:} ALMA $^{12}$CO(2--1) integrated intensity (moment~0) image of EX~Lup. \reply{Contours start at 3$\sigma$ with steps of 10$\sigma$ ($\sigma$=9~mJy\,beam$^{-1}$~km~s$^{-1}$).
{\it Center-panel:} Integrated intensity of $^{13}$CO(2--1). Contours start at 3$\sigma$ with steps of 5$\sigma$ ($\sigma$=7~mJy\,beam$^{-1}$~km~s$^{-1}$)}.
{\it Right-panel:} C$^{18}$O(2--1).  \reply{Contours start at 3$\sigma$ with steps of 1$\sigma$ ($\sigma$=5~mJy\,beam$^{-1}$~km~s$^{-1}$)}. The $^{12}$CO(2--1) data has been displayed using a {\tt sqrt} stretch in order to highlight faint extended emission.  \reply{The colorbar units are Jy\,beam$^{-1}$~km~s$^{-1}$. The star symbol in the center of each panel represents the stellar position, which coincides with the peak of the continuum emission. The velocity range used to make each map is annotated in the top right. The dashed white circle shows the region used to compute the integrated line emissions listed in Table~\ref{table-1}.}
}\label{fig-1}
\end{figure*}

Rotational line emission ($J$ = 2--1) from all three main
isotopologues of CO was detected. The vibrationally excited
$^{13}$CO(2--1, v = 1) line was not detected, \reply{and provides a
  1$\sigma$ upper limit of 4.7~mJy~km~s$^{-1}$ (assuming a line width
  of 9~km~s$^{-1}$)}. Figure~\ref{fig-1} shows the integrated emission
(moment~0) maps for each detected transition with continuum contours
overlaid. \reply{The $^{12}$CO integrated line emission is
  5011~$\pm$~510 mJy~km~s$^{-1}$ (integrated over circular aperture of
  1.7$\arcsec$ centered at the stellar position). The $^{13}$CO and
  C$^{18}$O total integrated emissions are 341~$\pm$~37 and
  77~$\pm$~10 mJy~km~s$^{-1}$ (integrated over circular apertures of
  0.75$\arcsec$ and 0.55~$\arcsec$, respectively). The quoted error
  includes both the statistical noise and absolute flux uncertainty,
  added in quadrature.}

The $^{12}$CO emitting region is notably more extended than the dust
continuum, $^{13}$CO, and C$^{18}$O emitting regions. \reply{The
  $^{12}$CO can be traced above 3$\sigma$ out to $\sim$~1.3$\arcsec$
  ($\sim$200~au) from the stellar position, while $^{13}$CO and
  C$^{18}$O extend out to $\sim$~0.5$\arcsec$ and $\sim$~0.3$\arcsec$
  respectively. We note that the full extension of the observed
  $^{12}$CO emission ( $\sim$3.4$\arcsec$ at 1$\sigma$) is close to
  the MRS allowed by the observations.}

\begin{table*}
    \caption{Continuum and Line Imaging Parameters}
  \label{table-1}
  \begin{center}
    \leavevmode
    \begin{tabular}{lccccc} \hline \hline                 
Source Properties      & 1.29~mm  &  1.37~mm & $^{12}$CO &  $^{13}$CO & C$^{18}$O  \\

\hline \hline   
Ra\tablenotemark{a}  & 16:03:05.479 & 0.000 & $-$0.015  & $-$0.013  & $-$0.014  \\
Dec\tablenotemark{a} & $-$40:18:25.789 &0.000 & +0.010  & +0.006    & +0.019 \\ 
Major axis FWHM ($\arcsec$)\tablenotemark{b}            &  0.45~$\pm$0.003  &  0.45~$\pm$0.007 &  - &  - & -  \\
Minor axis FWHM ($\arcsec$)\tablenotemark{b}            &  0.37~$\pm$0.002  &  0.38~$\pm$0.005 &  - &  - & -  \\
Position Angle ($^{\circ}$)\tablenotemark{c}         &  63.1~$\pm$1.2  & 62.1~$\pm$3.0  &  80~$\pm$2 & 76~$\pm$3 & 71~$\pm$4  \\
Peak Intensity\tablenotemark{d}    &  8.8   &  8.2  & 605.9 &  94.7 & 34.4 \\
Integrated  Flux\tablenotemark{e} & 17.37~$\pm$~0.15 & 15.16~$\pm$~0.32 & 5011~$\pm$~510 & 341~$\pm$~37 & 77~$\pm$~10 \\
Line Centroid \tablenotemark{f} (km~s$^{-1}$)    &  -     &  -    & 3.7~$\pm$~0.05 &  3.9~$\pm$~0.2 & 5.0~$\pm$~0.6   \\
\hline  \hline 
Beam Properties and image RMS & & & & \\ 
\hline   
Major axis ($\arcsec$)       &  0.32  &  0.34   &  0.39 &  0.41 &  0.41   \\
Minor axis ($\arcsec$)       &  0.26  &  0.28   &  0.29 &  0.31 &  0.32   \\
Position Angle ($^{\circ}$)   &  61.8  &  61.43  &  62.7 & 63.9  &  63.4   \\
RMS      (mJy\,beam$^{-1}$)   &  0.038 &   0.1   &   4.5 &  3.6 &    2.2    \\
RMS Moment 0 (mJy\,beam$^{-1}$~km~s$^{-1}$)&  - &   -   &   8.9 &  7.3 &    5.3     \\

\hline
    \end{tabular}
    \tablenotetext{a}{Centroid coordinates for 1.37~mm, $^{12}$CO,
      $^{13}$CO and C$^{18}$O  relative to the 1.29mm 
      centroid, in units of arcseconds.}
    \tablenotetext{b}{\resp{Convolved with the beam.}}

       \tablenotetext{c}{\reply{PA
        for continuum images derived from {\sc imfit}. PA for line
        images derived from spectro-astrometry. }}
    \tablenotetext{d}{Units are mJy\,beam$^{-1}$ for continuum and
      mJy~\,beam$^{-1}$~km~s$^{-1}$ for moment 0 images.}
      \tablenotetext{e}{Units are mJy for
      continuum and mJy~km~s$^{-1}$ for line images. Integrated fluxes
      for spectral line images are obtained by integrating over
      \reply{apertures of 1.7, 0.75 and 0.55$\arcsec$ for $^{12}$CO,
        $^{13}$CO and C$^{18}$O respectively.}
      \tablenotetext{f}{Derived
        from Gaussian fits to the line profiles.}
      }
  \end{center}
\end{table*}



\begin{figure*}
\centering\includegraphics[trim={1cm 2cm -1cm -1cm}, clip, width=\textwidth]{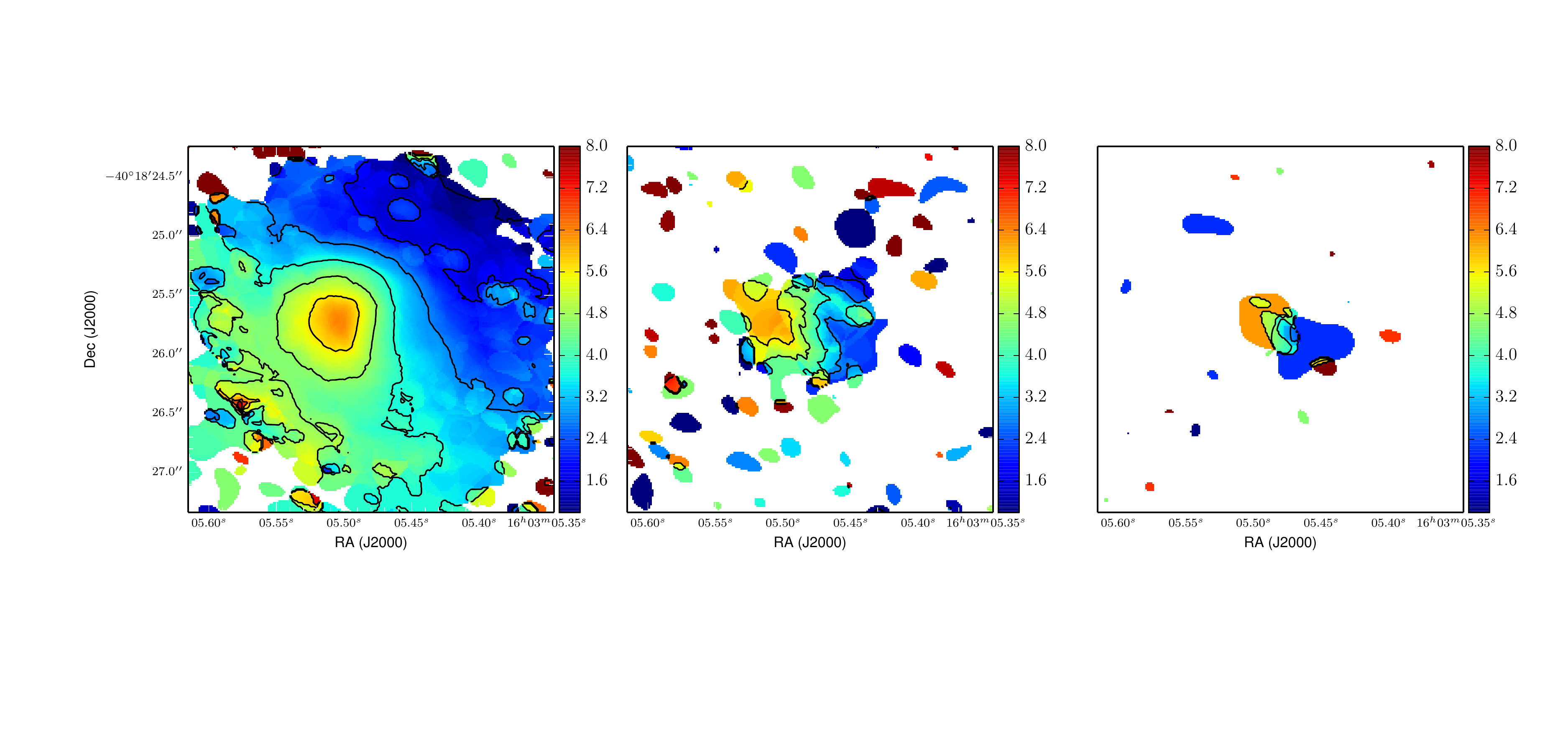}
\caption{{\it Left to right:} Moment~1 maps for $^{12}$CO(2--1), $^{13}$CO(2--1), and
  C$^{18}$O(2--1), respectively. Only pixel values above 3$\sigma$ have been
  included. Contour levels are 0.9, 1.8, 2.8, 3.7, 4.7, 5.6, 6.5, and
  7.5~km~s$^{-1}$.}
\label{fig-mom1}
\end{figure*}
\begin{figure*}
\includegraphics[angle=0,width=0.35\textwidth]{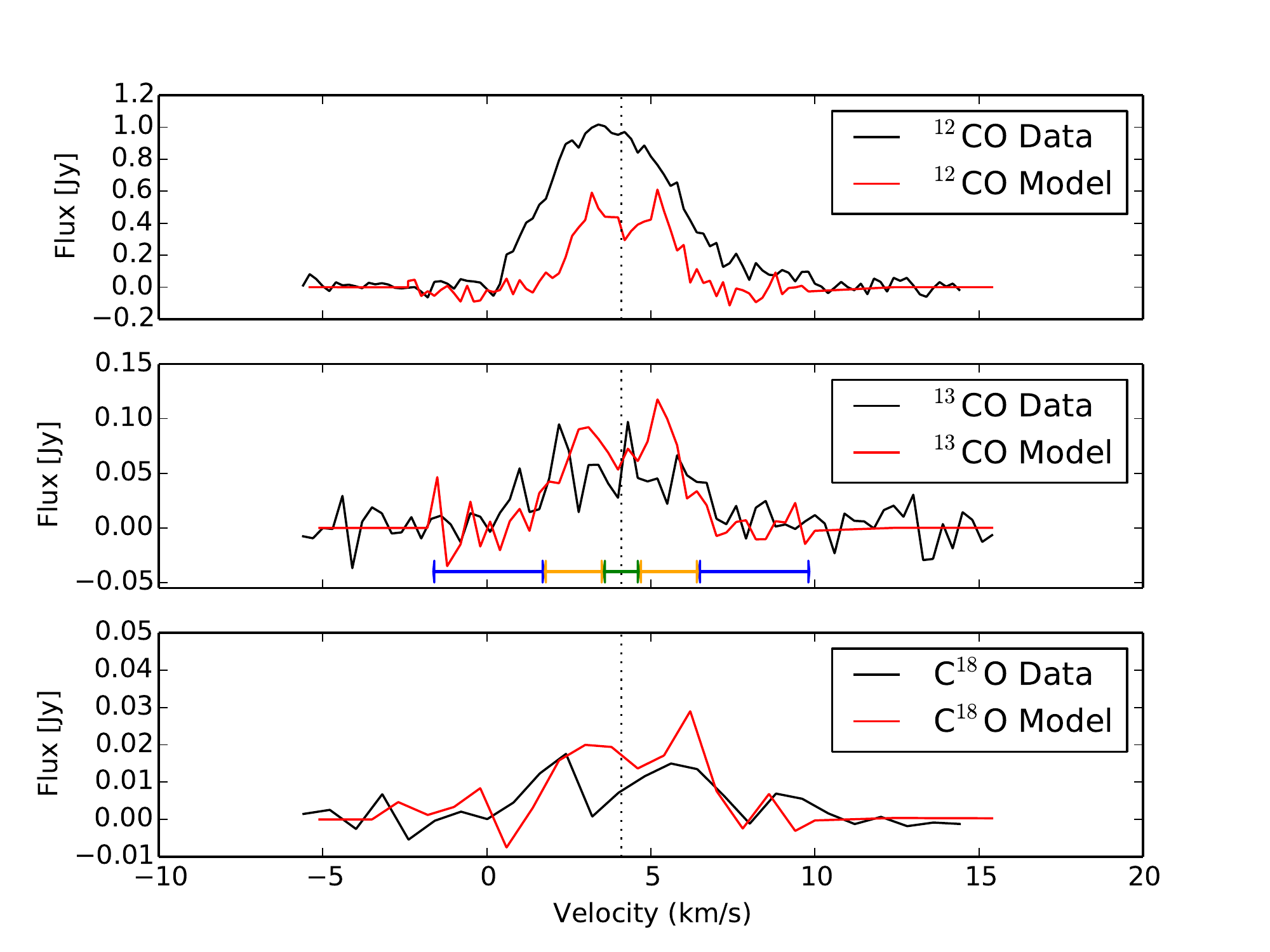}\hspace{-0.5cm}
\includegraphics[angle=0,width=0.35\textwidth]{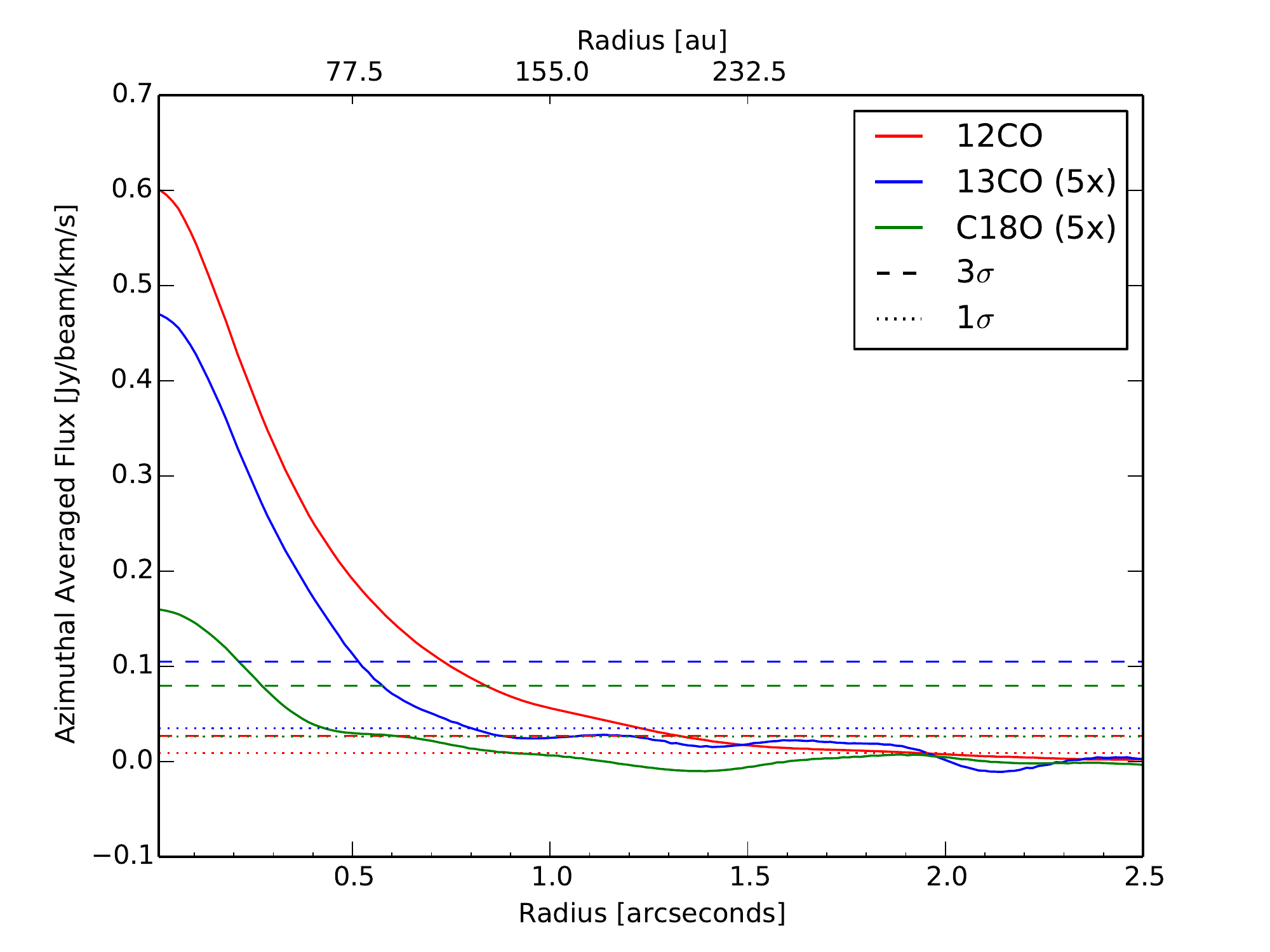}\hspace{-0.5cm}
\includegraphics[angle=0,width=0.35\textwidth]{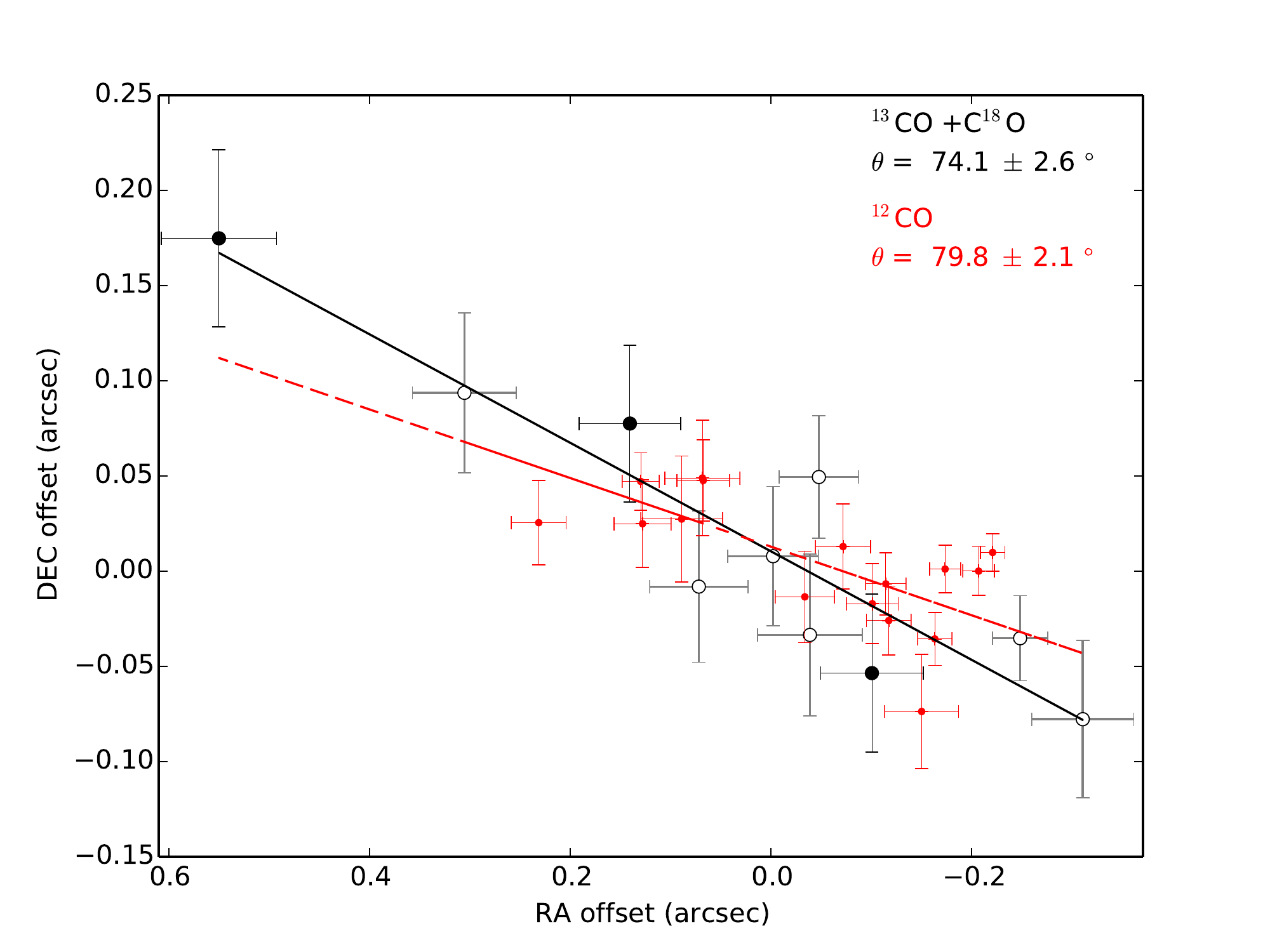}
\caption{ {\it{Left panel:}} Integrated spectrum for $^{12}$CO(2--1)
  (top), $^{13}$CO(2--1) (middle), and C$^{18}$O(2--1)
  (bottom). Spectra were computed by integrating the line images over
  \reply{apertures of 1.7, 0.75 and 0.55$\arcsec$ for $^{12}$CO,
    $^{13}$CO and C$^{18}$O respectively. \reply{The red lines show the
      spectra for the gas disk model presented in
      Section~\ref{co-models} }. The vertical dotted line represents the system's velocity, while the horizontal blue, orange and green lines represent the velocity ranges use to construct the moment~0 maps at different velocity  intervals from Figure~\ref{fig-12co-mom-vels}. {\it{Central panel:}} Azimuthally
    averaged surface brightness profiles for $^{12}$CO, $^{13}$CO and
    C$^{18}$O moment 0 images. The dashed and dotted lines represent
    the 3-$\sigma$ and 1-$\sigma$ levels of each map. $^{13}$CO and
    C$^{18}$O have been multiplied by a factor of 5 for display
    purposes.}  {\it{Right panel:}} Spectro-astrometry measurements of
  the $^{13}$CO(2--1) and C$^{18}$O(2--1) channel maps (open and
  filled black circles, respectively). Each data point corresponds to
  the centroid of the line emission relative to the center of the
  continuum emission. Only channels with emission above a SNR of 5
  have been included. The black line corresponds to a ${\chi}^2$ fit
  to the combined $^{13}$CO(2--1) and C$^{18}$O(2--1) data. The
  resulting position angle is annotated in the image, together with
  the derived uncertainty from the fit. The error bars correspond to
  the astrometric accuracy of the data. Red points show the
  spectro-astrometry measurements of the $^{12}$CO(2--1) high velocity
  channel maps relative to the center of the continuum
  emission. Velocity channels between +0.6 and +1.6~km~s$^{-1}$ were
  selected for the blue side, while channels between +6.8 and
  +8.6~km~s$^{-1}$ were selected for the red side. The red dashed line
  shows the best fit to the $^{12}$CO(2--1) data.  }
\label{fig-astro}
\end{figure*}

Figures~\ref{fig-mom1} and ~\ref{fig-astro} show the
intensity-weighted velocity fields (moment~1) and the integrated
spectrum for the three CO isotopologues. The moment~1 images are
indicative of rotation around the continuum disk, although only
$^{13}$CO(2--1) and C$^{18}$O(2--1) seem to trace mostly Keplerian
rotation at a position angle of approximately
70$^{\circ}$--80$^{\circ}$. We performed astrometry of the channel
maps of $^{13}$CO and C$^{18}$O (presented in Appendix~\ref{co_maps})
in order to determine the position angle of the disk by measuring the
centroid position of the red-shifted and blue-shifted emission of each
line independently. Only channels where the emission has an SNR~$>$~5
were included in the fit. This resulted in 7 data points for $^{13}$CO
and 3 data points for C$^{18}$O. All data points align well with the
peak of the continuum (Fig.~\ref{fig-astro}, right panel). By
performing a ${\chi}^2$ fit we determined the PA to be
76$^{\circ}$~$\pm~$3$^{\circ}$ and 71$^{\circ}$$~\pm~$3$^{\circ}$ for
$^{13}$CO and C$^{18}$O respectively. Fitting both the $^{13}$CO and
C$^{18}$O data simultaneously yielded a PA =
74$^{\circ}$$~\pm~$3$^{\circ}$.

\begin{figure*}
\centering\includegraphics[width=0.9\textwidth]{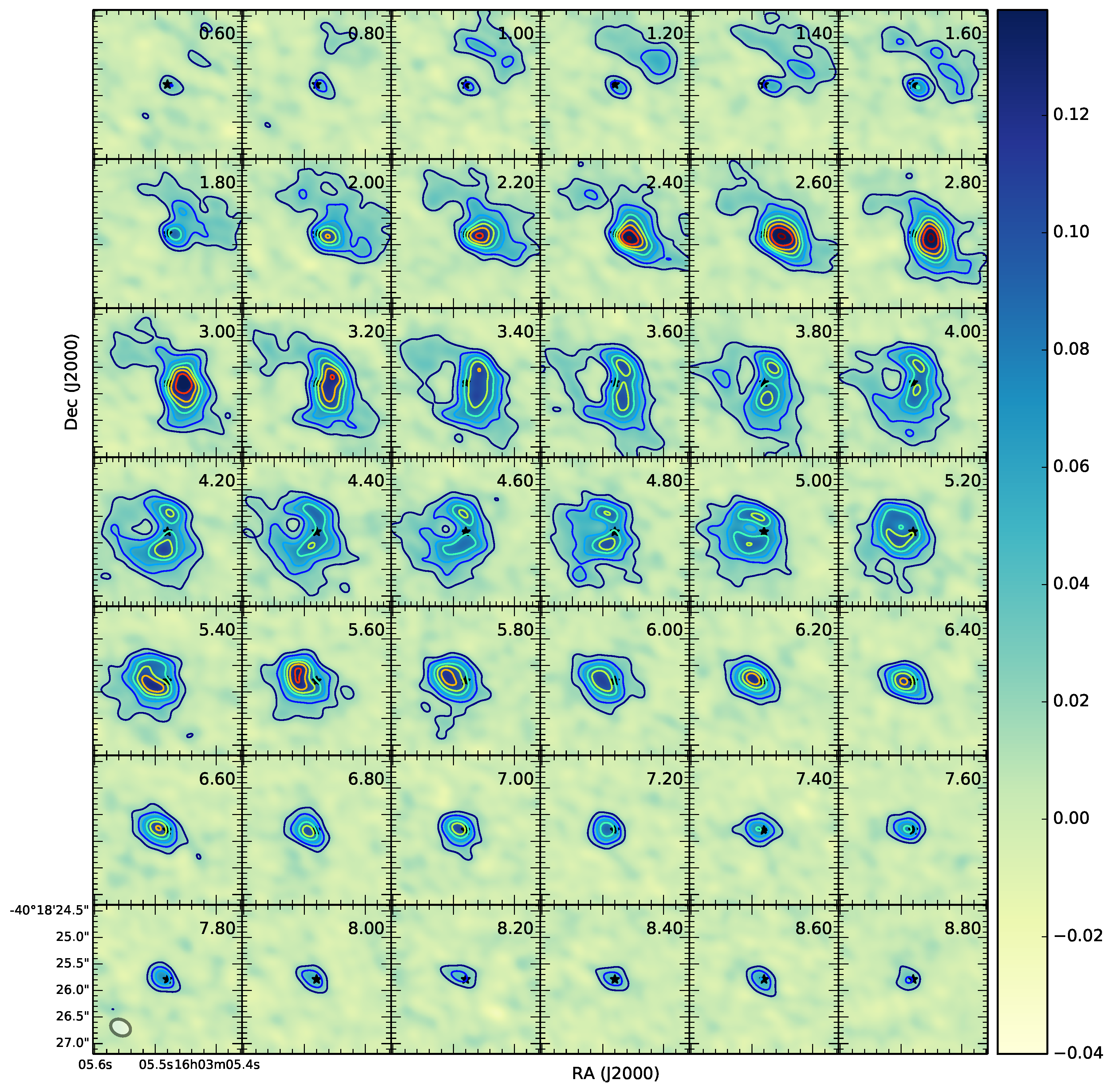}
\caption{ $^{12}$CO channel maps towards EX~Lup. The velocity of the
  channels is shown in the Local Standard of Rest (LSR) frame,
  centered at the rest frequency of $^{12}$CO(2--1). The data has
  been binned to a velocity resolution of 0.2~km~s$^{-1}$. Contours of the $^{12}$CO
  emission are overlaid. Contour
  levels are
  0.018, 0.036, 0.054, 0.072, 0.09, 0.11, 0.13, and 0.15~Jy\,beam$^{-1}$. The star
  symbol in the center of each panel
  represents the stellar position, coinciding with the peak of
  the continuum emission.}\label{fig2_12co}
\end{figure*}

Like the $^{13}$CO and C$^{18}$O first moments, the $^{12}$CO first
moment (Fig.~\ref{fig-mom1}) has a velocity gradient along the disk
major axis typical of disk rotation, although with deviations from a
pure Keplerian pattern (and at a scale larger than for the $^{13}$CO
and C$^{18}$O emission). The $^{12}$CO first moment also shows a
velocity gradient perpendicular to the disk's major axis, consistent
with the presence of an outflow.  The $^{12}$CO channel maps
(Fig.~\ref{fig2_12co}) reveal an asymmetric, rotating structure, near
a systemic velocity somewhere between +3.6 and +4.6~km~s$^{-1}$. There
is a ``twist'' at systemic velocities seen in the $^{12}$CO first
moment image (Fig.~\ref{fig-mom1}). This is similar to the
infalling-rotating motion from a tenuous envelope into a disk observed
in Class 0/I sources \citep[e.g.][]{yen2014,oya2016}, and recently
discovered in Class II sources \citep{2017arXiv170802384Y}.  The red
side of the disk (v$~>~$+4.4~km~s$^{-1}$) exhibits the ``butterfly''
pattern typical of a rotating Keplerian disk seen at intermediate
inclination. This is in agreement with previous works that have
estimated the disk inclination angle as between about 20$^{\circ}$ and
50$^{\circ}$ \citep{sipos2009,goto2011,kospal2014,Sicilia2015} and is
consistent with our estimate of 32$^{\circ}$ from the continuum
image. Optical absorption-line measurements \citep{kospal2014} suggest
an optical systemic LSR velocity of $\sim~$+4.4~km~s$^{-1}$, in
reasonable agreement with centroid velocities of the CO line profiles,
particularly for the more optically-thin and spatially-compact
$^{13}$CO and C$^{18}$O lines (Table~\ref{table-1}). \reply{The line
  centroids determined from gaussian fitting (Table~\ref{table-1}) are
  virtually identical to the line positions measured by
  \citet{kospal2016b} using single-dish APEX spectra (3.7, 3.9 and
  5.0~km~s$^{-1}$ for ALMA versus 3.6, 3.9 and 5.1~km~s$^{-1}$ for
  APEX).}

\begin{figure*}
\includegraphics[scale=0.27]{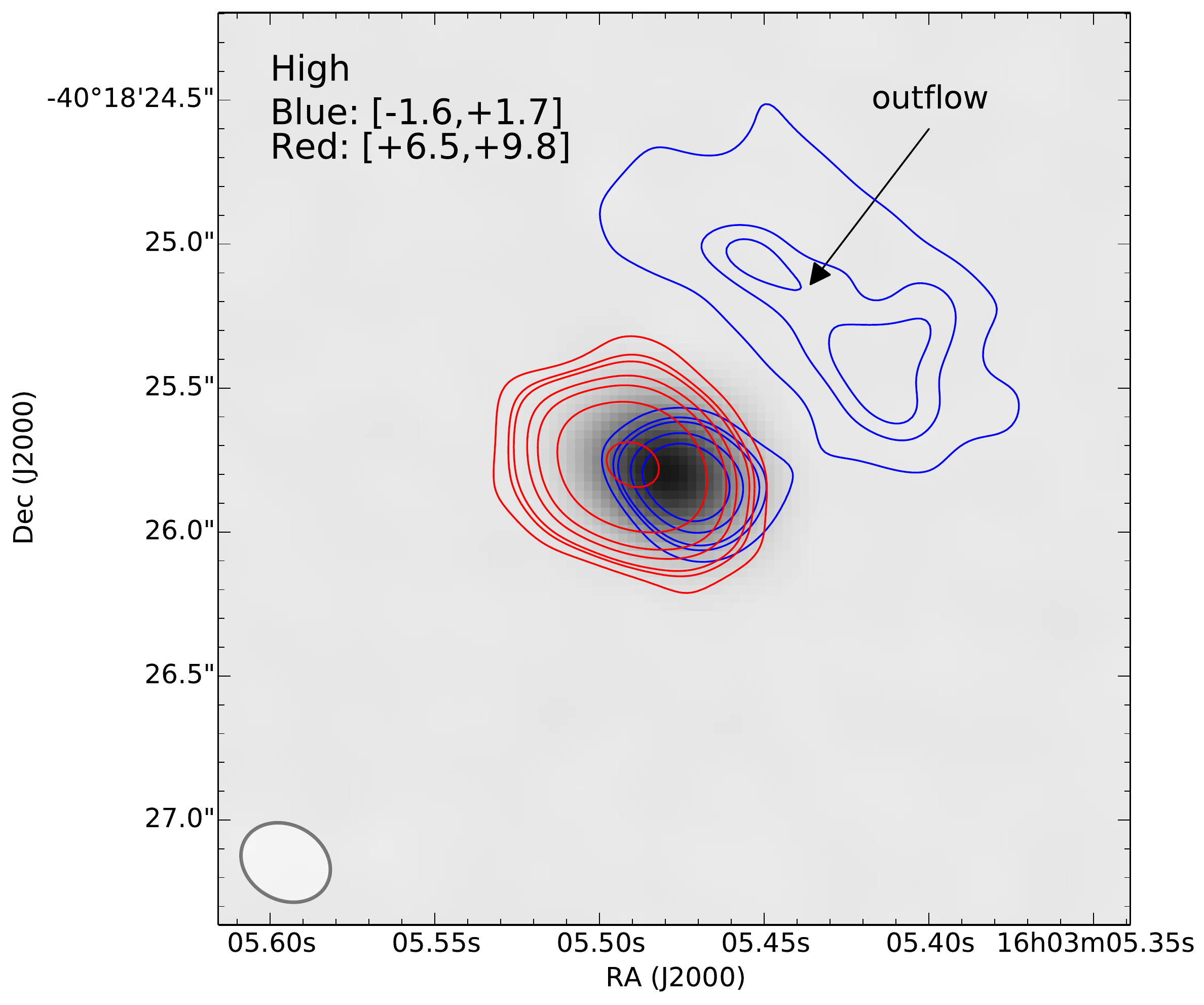}
\includegraphics[scale=0.27]{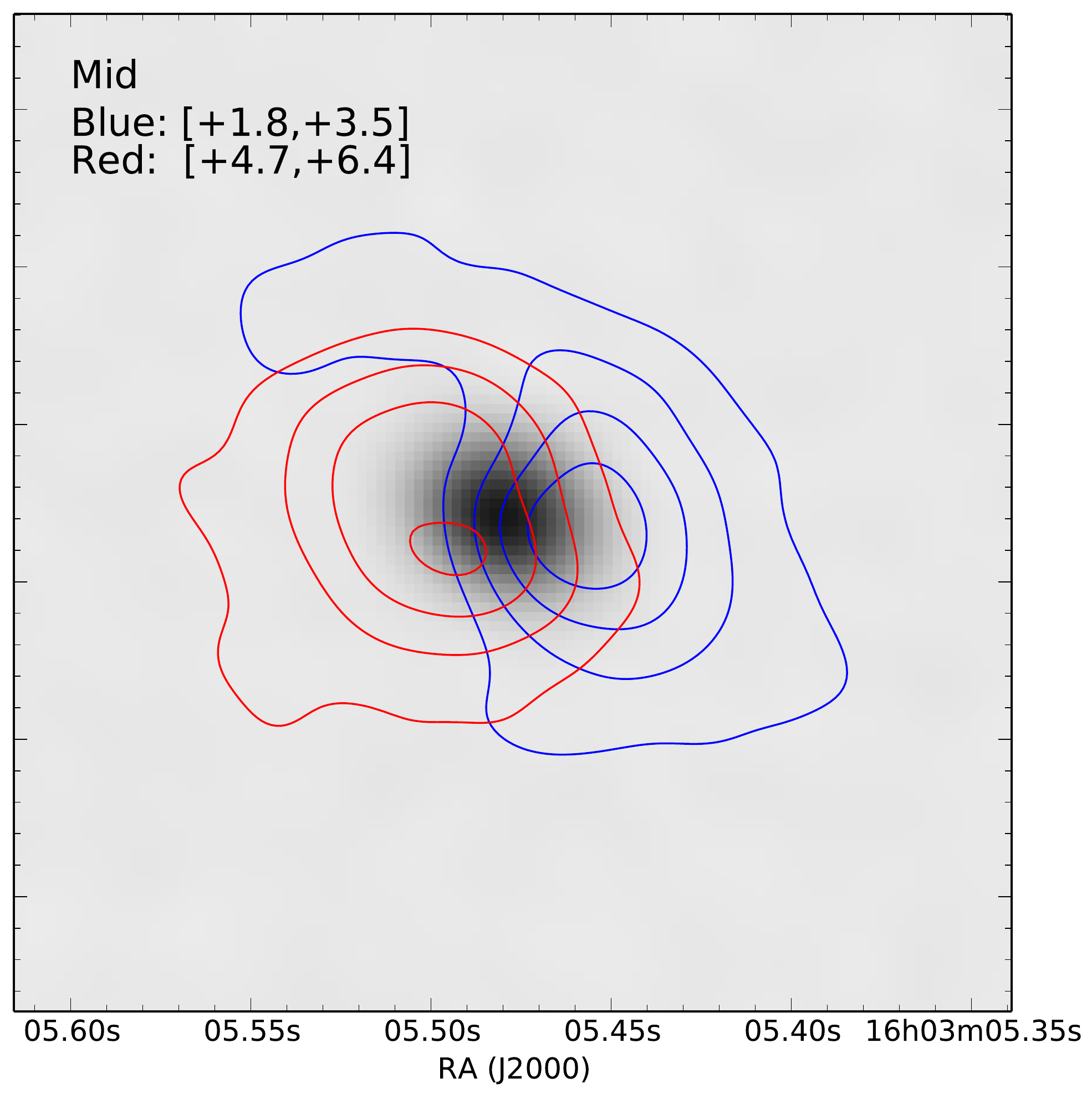}
\includegraphics[scale=0.27]{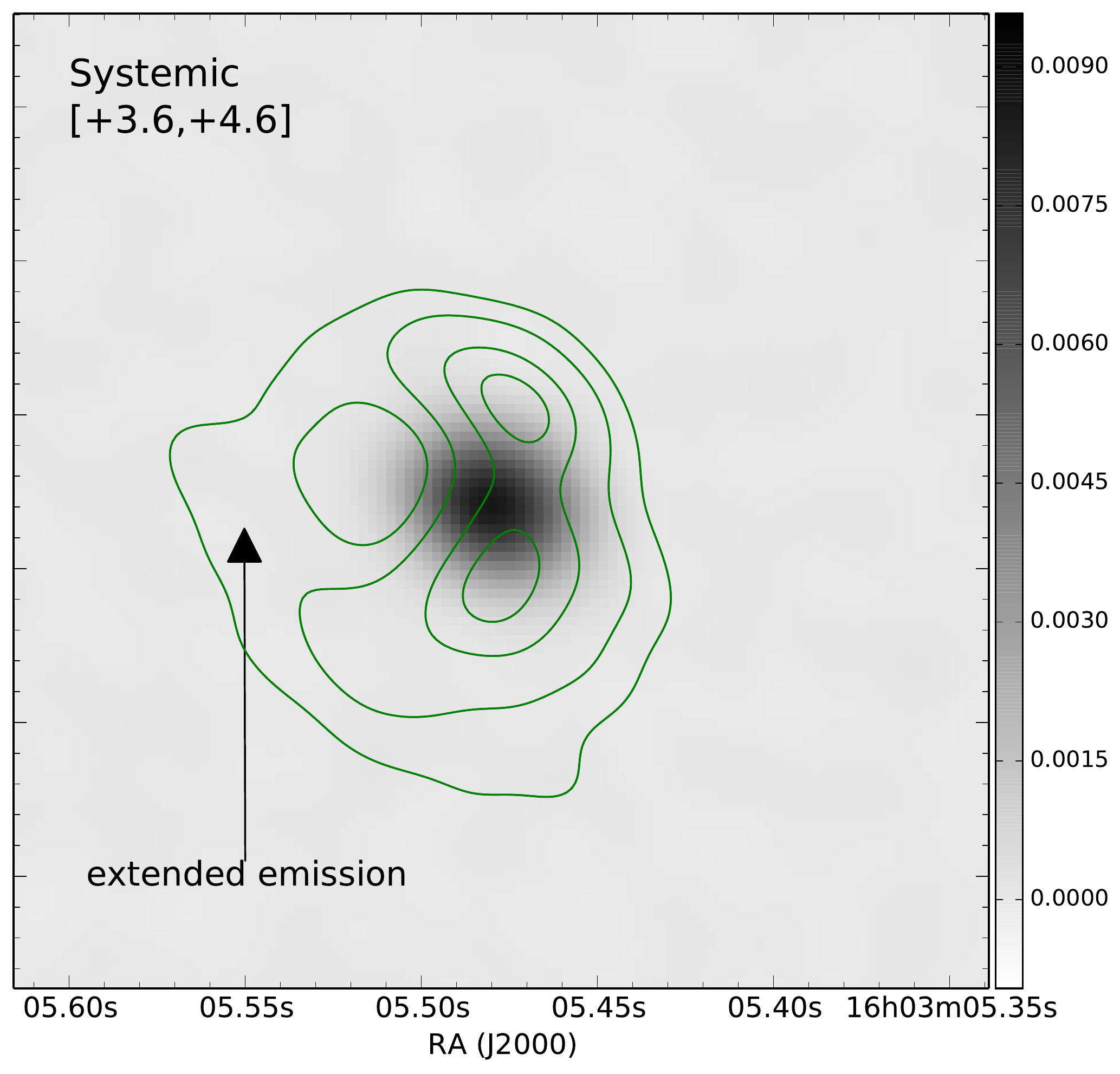}
\caption{$^{12}$CO(2--1) moment~0 maps at different velocity ranges.
  \reply{The high velocity channels ({\it{Left-panel}}) have been
    integrated between -1.6 and +1.7~km~s$^{-1}$ (blue) and between
    +6.5 and +9.8~km~s$^{-1}$ (red).  \resp{The grayscale shows 1.29mm
    continuum emission from the compact dust disk (in units of
    Jy\,beam$^{-1}$)}. Intermediate velocity channels
    ({\it{Center-panel}}) have been integrated between +1.8 and
    +3.5~km~s$^{-1}$ (blue) and between +4.7 and +6.4~km~s$^{-1}$
    (red). {\it Right-panel:} Integrated emission at systemic
    velocities (+3.6--+4.6~km~s$^{-1}$). Contour levels are same as
    for Figure~\ref{fig2_12co}.}}
\label{fig-12co-mom-vels}
\end{figure*}

To further highlight the differences between the blue and red side of
the $^{12}$CO(2--1) emission, Figure~\ref{fig-12co-mom-vels} shows the
integrated intensity maps at systemic, intermediate, and high velocity
ranges. For comparison, moment~0 maps for $^{13}$CO(2--1) and
C$^{18}$O(2--1) in identical velocity ranges are presented in
Appendix~\ref{co_maps}. The blue side of the disk is strikingly
different from the red side. Instead of mirroring the red-shifted
side, the blue-shifted emission is dominated by a bright, compact region at
systemic and intermediate velocities. The blue side shows an
additional extended spiral-shaped component shown in
Figure~\ref{fig-12co-mom-vels} up to 1.2$\arcsec$ in extent, and also
visible in the +2.6 and +2.8~km~s$^{-1}$ channel maps. At higher
blue-shifted velocities ($v~<~$+1.6~km~s$^{-1}$) there is emission near the central
star as a counterpart to the high velocity emission seen on the red
side. Another feature of the line maps is an extended component at
systemic and intermediate red-shifted velocities, extending up to
1.5$\arcsec$ from the star.

The compact high velocity emission on the blue side is fainter than
its red-shifted counterpart, and the red-shifted emission extends
to higher relative velocities than does the blue-shifted counterpart
(assuming a systemic velocity of $\sim~$+4.1~km~s$^{-1}$). The red-shifted
high velocity gas, which is located very close to the central star, can be traced
down to +9.4~km~s$^{-1}$, whereas the blue-shifted
compact emission is detected only out to +0.2~km~s$^{-1}$. The overall
rotation follows a slightly larger PA than derived for $^{13}$CO and
C$^{18}$O. We performed a similar astrometric analysis for
$^{12}$CO(2--1) as for the two isotopologues, but selecting high
velocity channels between +0.6 and +1.6~km~s$^{-1}$ for the blue side
and between +6.8 and +8.6~km~s$^{-1}$ for the red side. This resulted in
a PA of 80$^{\circ}~\pm~$2$^{\circ}$ (Fig.~\ref{fig-astro}).

Between +1.2 and +1.6~km~s$^{-1}$, we note extended, arc-like
emission located north-west of the star. It is located 0.8$\arcsec$
(125~au) away from the disk plane, in the direction of the rotational
axis of the system where molecular outflows are typically seen. We
interpret this arc-like feature as the interaction of a molecular outflow with remnant
ambient material. Such arc-like shapes have been discovered
recently in other outbursting sources
\citep{ruiz2017,zurlo2017,kospal2017b,principe2018}, although mostly
around FUors and one transitional FUor/EXor source \citep{principe2018}. 

Given the strong asymmetry between the red-shifted and blue-shifted
sides, determining the exact systemic velocity is difficult. The typical
``hourglass'' shape from an inclined Keplerian disk seen at the
systemic velocity is normally used to identify the central velocity
using channel maps. In the EX~Lup channel maps, channels between +3.6 and
+4.4~km~s$^{-1}$ could plausibly be attributed to the systemic velocity, with
+3.6~km~s$^{-1}$ perhaps slightly favored. However not only is the hourglass shape
slightly twisted, but the center of the hourglass is not located at the position of
the continuum as is typical of circumstellar disks (see
section~\ref{asymmetry}). The double-lobed pattern only becomes centered with the
continuum centroid at velocities of +4.0 to +4.2~km~s$^{-1}$, which is closer to the
optically-measured system velocity. The $^{12}$CO spectrum
(Fig.~\ref{fig-astro}) does not exhibit a double peaked profile as
expected from a purely rotating, inclined, symmetric disk. Instead, it
shows a single peaked profile centered at +3.7~km~s$^{-1}$. The
$^{12}$CO(2--1) line shape is similar to the $^{12}$CO(3--2) line
profile from APEX single-dish observations \citep{kospal2016b}, which also
peaks at +3.6~km~s$^{-1}$.

\begin{center}
\begin{figure*}
\includegraphics[width=0.45\textwidth]{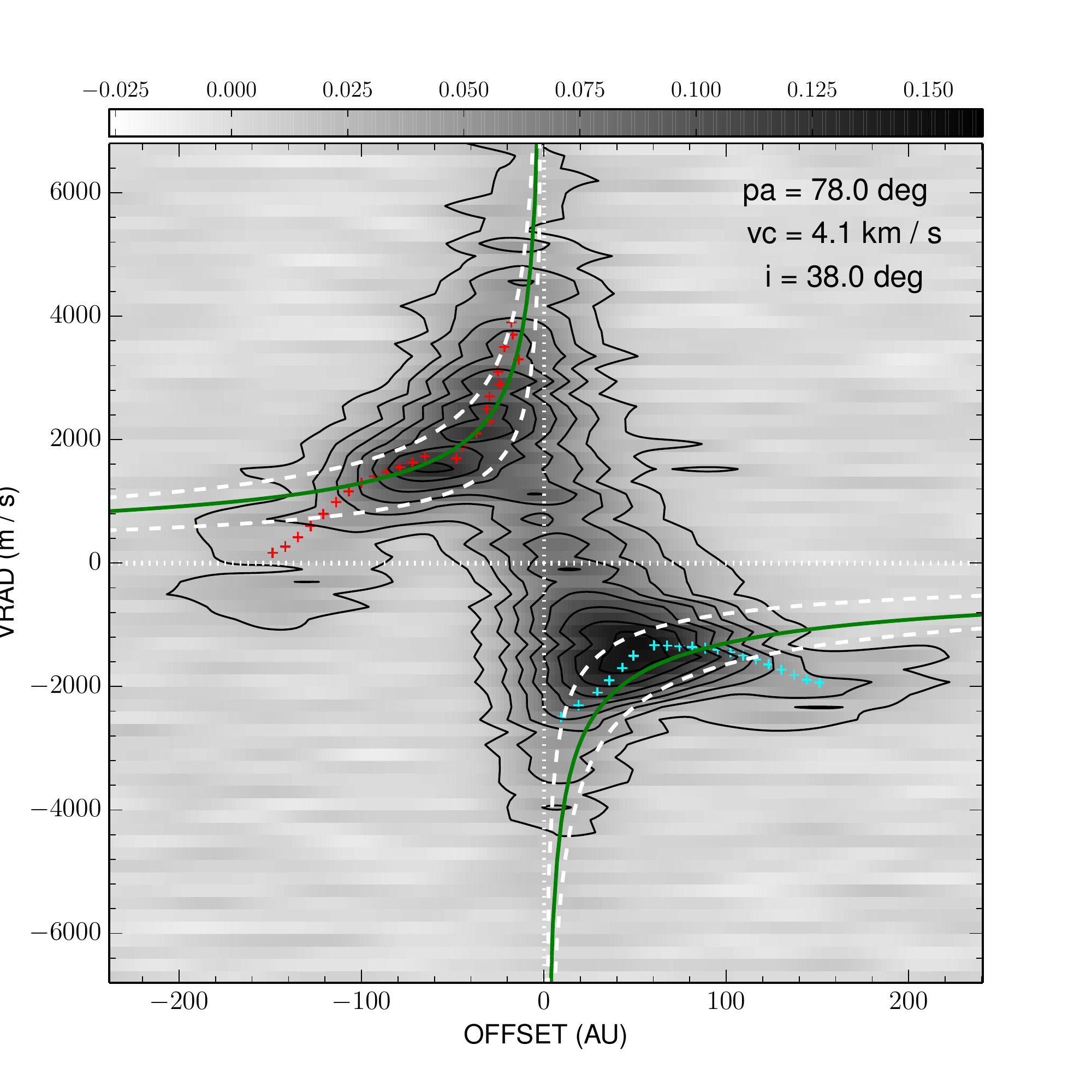}
\includegraphics[width=0.59\textwidth]{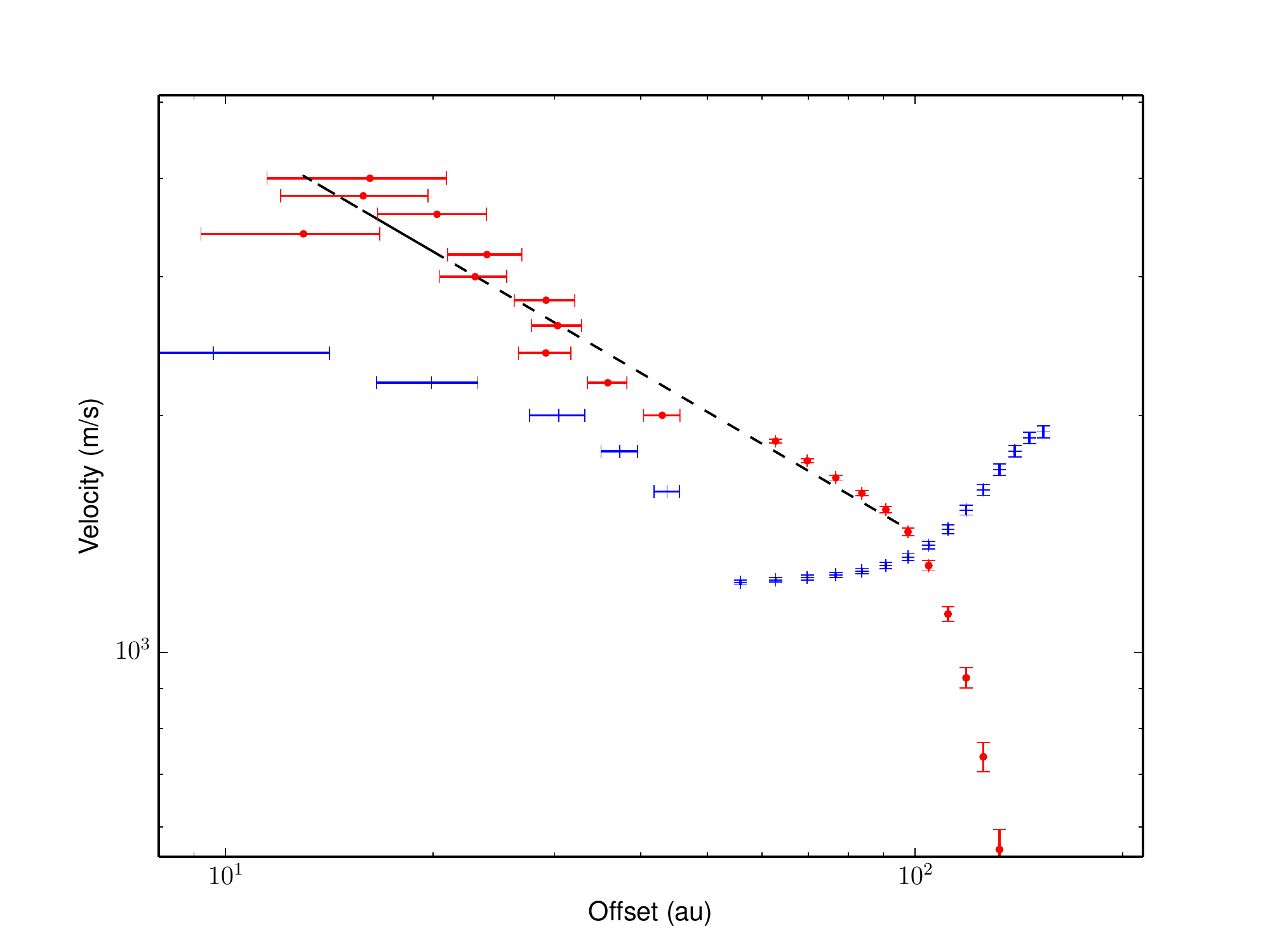}
\caption{ {\it{Left-panel:}} $^{12}$CO position-velocity diagram
  along the disk major axis (assumed PA =
  78$^{\circ}$). The horizontal dashed white line represents the possible systemic
  velocity of +4.1~km~s$^{-1}$, while the vertical dotted black line represents the
  stellar position. The red and blue markers show measured
  centroid positions in different velocity channels for the red-
  and blue-shifted sides of the disk. The green curves show the
  Keplerian velocity for gas at the tangential point at an inclination
  angle of 38$^{\circ}$. \reply{The mass of the central object was
  assumed to be 0.5~M$_{\odot}$ \citep{frasca2017}}. For comparison, the
  Keplerian velocities for a 0.2 and a 0.8~M$_{\odot}$ central star are also shown
  (inner and outer dashed white lines respectively).
  {\it{Right-panel:}} Rotation profile of the $^{12}$CO red-shifted
  (red) and blue-shifted (blue) emission. The dashed line represents
  the results from least-square fitting to the inner $r~<~$100~au
  red-shifted data points, which results in an $r^{-0.53}$ slope. The
  blue-shifted emission contrarily has a positive velocity gradient at
  $r~>~$50~au.}\label{fig-pv-co}
\end{figure*}
\end{center}

\begin{figure}
\centering\includegraphics[scale=0.4]{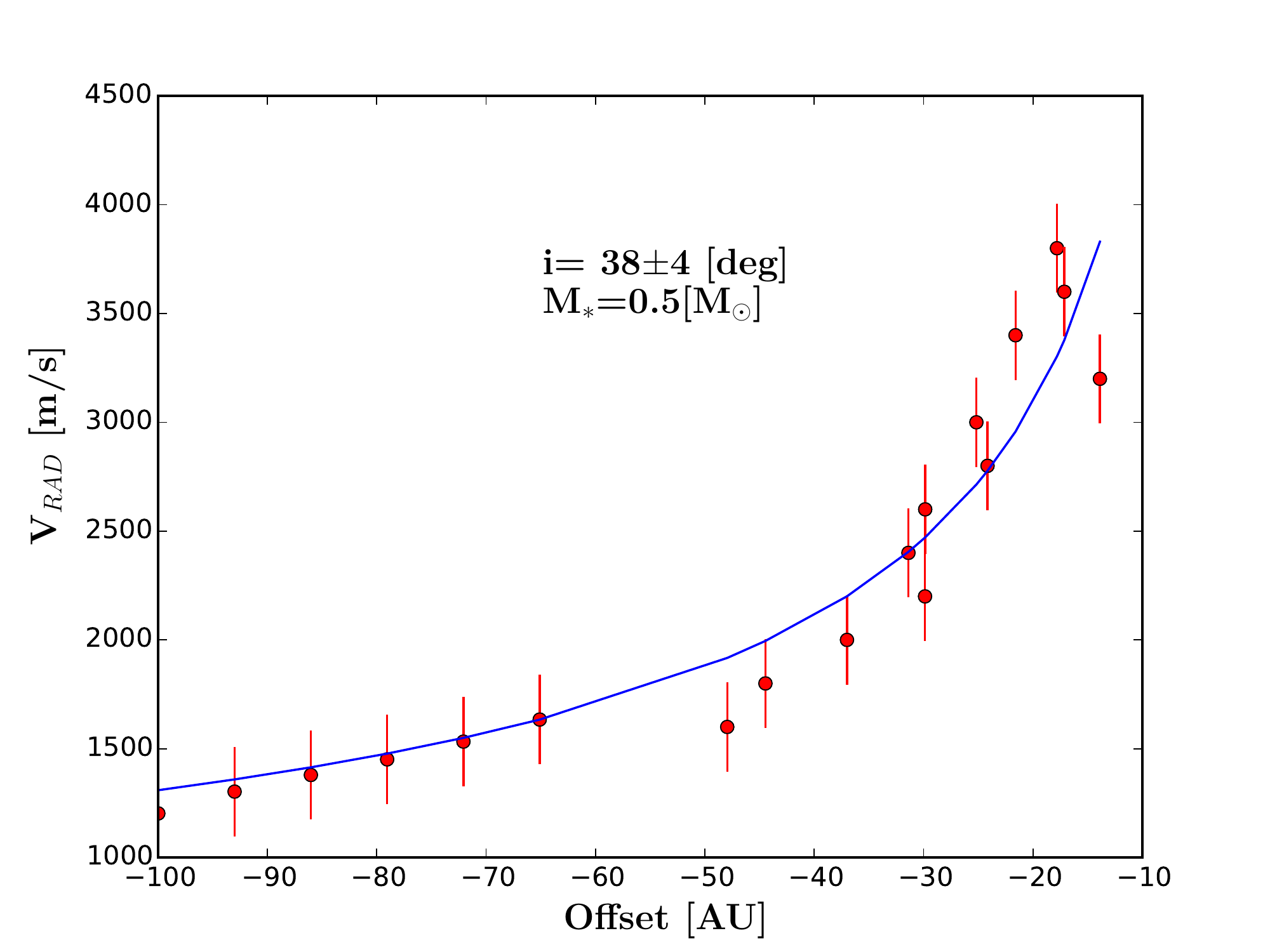}\hspace{-1.cm}
\caption{Best fit to the inclination of angle to the inner $r~<~$100~au
  red-shifted data points from Figure~\ref{fig-pv-co}. \reply{The stellar
  mass was assumed a value of 0.5~M$_{\odot}$}.
}\label{fig-angle}
\end{figure}

Position-velocity (PV) diagrams for all three isotopologues were
constructed by extracting the averaged spectra along the disk major
axis (assuming a PA of 78$^{\circ}$). The features of the
$^{12}$CO(2--1) maps are evident in the PV diagram
(Fig.~\ref{fig-pv-co}): the red-shifted emission is slightly fainter
than the blue, and shows compact high velocity emission near the
star. The blue-shifted emission from part of the high velocity
arc-like feature is evident at $r~>~$100--200~au. The emission on the
red-shifted side seems consistent with a Keplerian disk of inclination
close to 40$^{\circ}$ (i.e. the value adopted by \citet{kospal2016b}
for modelling the single-dish CO spectra). In Figure~\ref{fig-pv-co},
we extract the centroid velocities as determined from a Gaussian fit
to the $^{12}$CO(2-1) line profiles on both sides of the disk and plot
them against position. The velocities from the red-shifted side can be
fit with an $r^{-0.53}$ power-law, consistent with Keplerian
rotation. However, the blue-shifted velocities deviate significantly
from Keplerian rotation, and increase with radius at $r~>~$55~au. In
Figure~\ref{fig-angle} we use the red-shifted data points from the
inner $r~<~$100~au from Figure~\ref{fig-pv-co} to fit the inclination
angle, assuming Keplerian rotation around a \reply{ star of mass
  0.5~M$_{\odot}$}. The inclination angle derived from ${\chi}^2$
fitting is 38$^{\circ}~\pm~$4, thus consistent with the value derived
by \citet{kospal2016b}.

\begin{figure*}
\includegraphics[width=0.5\textwidth]{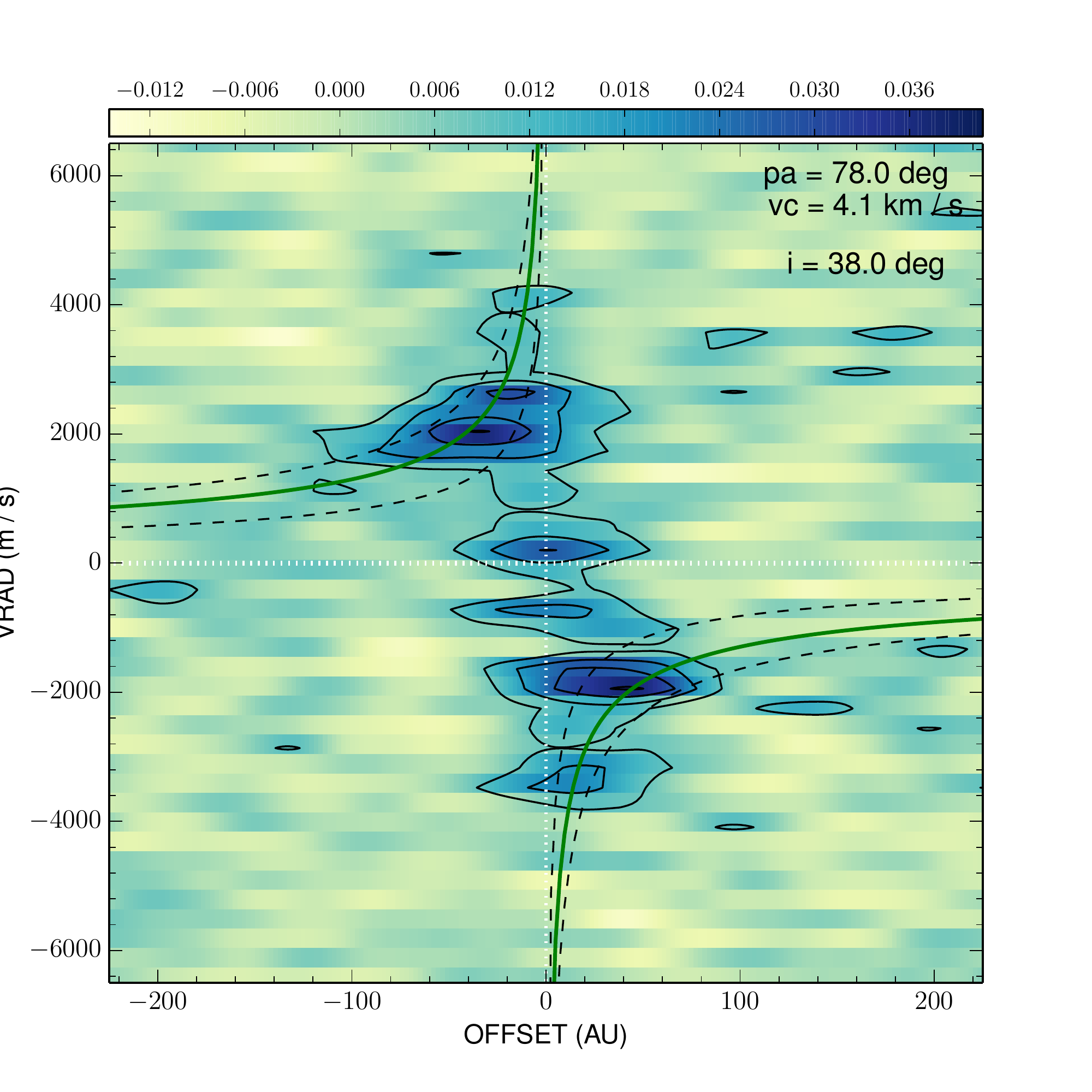}
\includegraphics[width=0.5\textwidth]{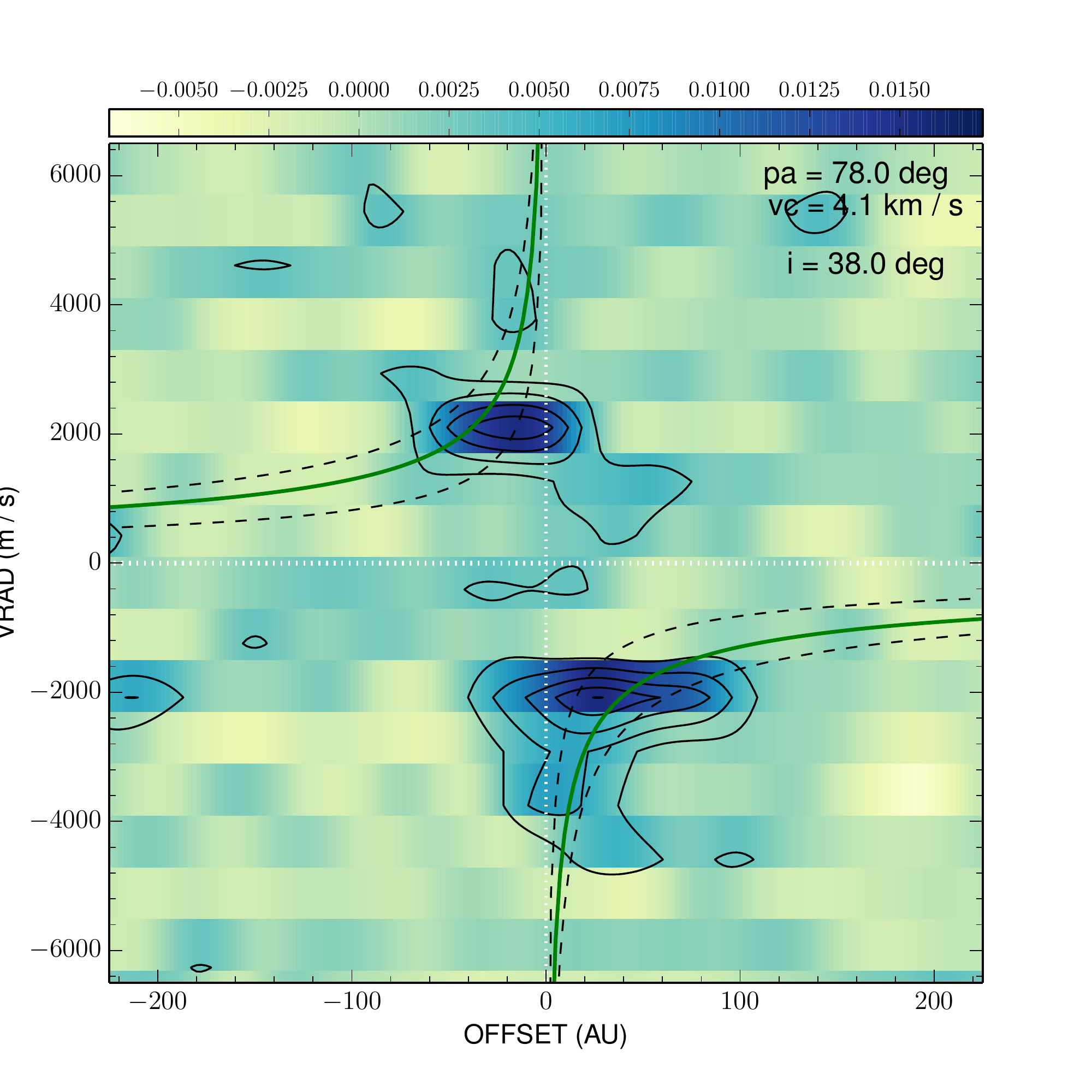}  
\caption{Position-velocity diagram for the observed $^{13}$CO ({\it
    Left-panel}) and C$^{18}$O ({\it Right-panel}) along the disk
  major axis (assuming a PA of 78$^{\circ}$). The horizontal dashed
  white line represents the possible systemic velocity of
  +4.1~km~s$^{-1}$, while the vertical dotted black line represents
  the stellar position. The green curves show the Keplerian velocity
  for gas at the tangential point at inclination angle of 38$^{\circ}$
  and a \reply{stellar mass of 0.5~M$_{\odot}$}. The Keplerian
  velocities for a 0.2 and a 0.8~M$_{\odot}$ central star are also
  shown (inner and outer dashed black lines
  respectively).}\label{fig-pv13-18-co}
\end{figure*}

 
The $^{13}$CO and C$^{18}$O data do not show any clear asymmetries in the
PV diagrams (Fig.~\ref{fig-pv13-18-co}) and are consistent with a
Keplerian disk with a systemic velocity between 
+4.0 and +4.4~km~s$^{-1}$. There is no evidence of outflows/inflows or additional
emission from the arc feature in $^{13}$CO or C$^{18}$O, suggesting that the arc CO
gas has a relatively low column density compared to the rest of the disk.

In the following section we use continuum and line radiative transfer
modeling to characterize the physical properties of the EX~Lup disk.

%

\section{Disk Modeling}\label{model}
\subsection{Radiative Transfer Modelling of Dust Continuum}\label{models}

The EX~Lup continuum emission presented here has no resolved
substructures and can be approximated as a continuous axisymmetric
disk. In order to derive disk structural parameters and the dust mass
the continuum was modeled using the radiative transfer code {\sc
  radmc-3d} \citep{Dullemond2012}. We adopt the standard tapered-edge
model to describe the surface density profile
\citep{andrews2009,williams11} parameterized by the total dust mass
$M_{\rm d}$, a power-law slope $\gamma$, a characteristic radius
$R_{\rm c}$, and a surface density $\Sigma$ that is tapered with an
exponential decay beyond $R_{\rm c}$:

\begin{eqnarray}
  \Sigma=\Sigma_{\rm c} \left( \frac{R}{R_{\rm c}} \right)^{-\gamma} \exp \left[- \left( \frac{R}{R_{\rm c}}\right)^{2-\gamma} \right] 
\end{eqnarray}\label{tapered}

The vertical scale height of the disk as a function of radius is also
a power law $H(r) = H_{100}(r/100\,{\rm au})^{\psi}$, with a flaring
index $\psi$~=~1.09 as determined by \citet{sipos2009}. A stellar spectrum with
$T_{\rm star}=3859$~K and a radius of 0.8~R$_\odot$ were adopted
\citep{frasca2017}. The model includes a standard power-law
distribution of dust grain sizes $a$, given by $n(a) \propto a^{-3.5}$ and
extending from $a_{\rm min} = 0.1\,\mu$m to $a_{\rm max} = 3.0\,$mm.
The optical properties of the dust correspond to a mix of amorphous
carbon grains from \citep{li1997} and astro-silicate grains \citep{draine1984}
which were combined using Bruggeman's rules.
We used a standard mix of 30$\%$ amorphous carbons and 70$\%$ astro-silicates. The
opacities of the mix were computed using the ``Mie Theory'' code written by
\citet{Bohren1983}. The mass opacity at 1.3~mm is thus
$\kappa_{\rm abs} = 2.2\,{\rm cm}^2\,{\rm g}^{-1}$.

The inner radius was set to approximately the expected dust
sublimation radius of $\sim$~0.05~au. This is smaller than the value
used by \citet{sipos2009}, who found that an inner radius smaller than
0.2~au caused an excess emission in the near- and mid-infrared part of
the SED. We caution that this radius depends on disk parameters for an
optically thick disk, but our current data does not constrain this
radius. Figure~\ref{fig-0} shows that our model fits the resolved
image of the disk at 1.3~mm, and it can be reconciled with the SED at
shorter wavelengths if the small and large grains follow different
spatial distributions \citep[as also noted by ][]{Sicilia2015}. Large
grains accumulate in the inner regions \citep[as predicted by radial
drift models, e.g. ][]{pinte2014} while the inner $<$~0.2~au region
is depleted of smaller grains.

\subsubsection{Fitting Procedure}

The model parameters that we aim to determine are $M_{\rm d}$,
$\gamma$, $R_{\rm c}$, and $H_{100}$. Before exploring the parameter space, the
centroid shift, the inclination angle $i$, and the PA of the continuum
emission were determined using the CASA task {\sc imfit}. The disk
structural parameters parameters were then constrained using a Bayesian
approach. The posterior distribution for each parameter was recovered using
the \citet{goodman2010} affine-invariant MCMC ensemble sampler as implemented by
\citet{foreman2013}. We used the latter's publicly available {\sc python} module
{\sc emcee} to sample the parameter space and
maximize the likelihood function. The likelihood function is
proportional to $\exp(-\chi^2/2)$, where $\chi^2$ is the sum over the
squared difference of the model and measured visibilities divided by
their variance. The model and the data (the self-calibrated Measurement
Set including only the continuum) were compared in the $uv$ plane. The
model visibilities were obtained by taking the Fast Fourier Transform
of model images and interpolating to the same $uv$ points as the observations \citep{marino}. 

Our priors for the free parameters came from assuming
uniform distributions, specifically:
\begin{eqnarray*}
  M_{\rm d} &\in& [0.001, 0.1] \times 10^{-2}\, {\rm M}_\odot\\
  \gamma &\in& [-2.0, 2.0]\\
  R_{\rm c} &\in& [5.0, 100.0]\, {\rm au}\\
  H_{\rm 100} &\in& [0.1, 10.0]\, {\rm au}
\end{eqnarray*}

\begin{figure*}
\centering\includegraphics[width=0.7\textwidth]{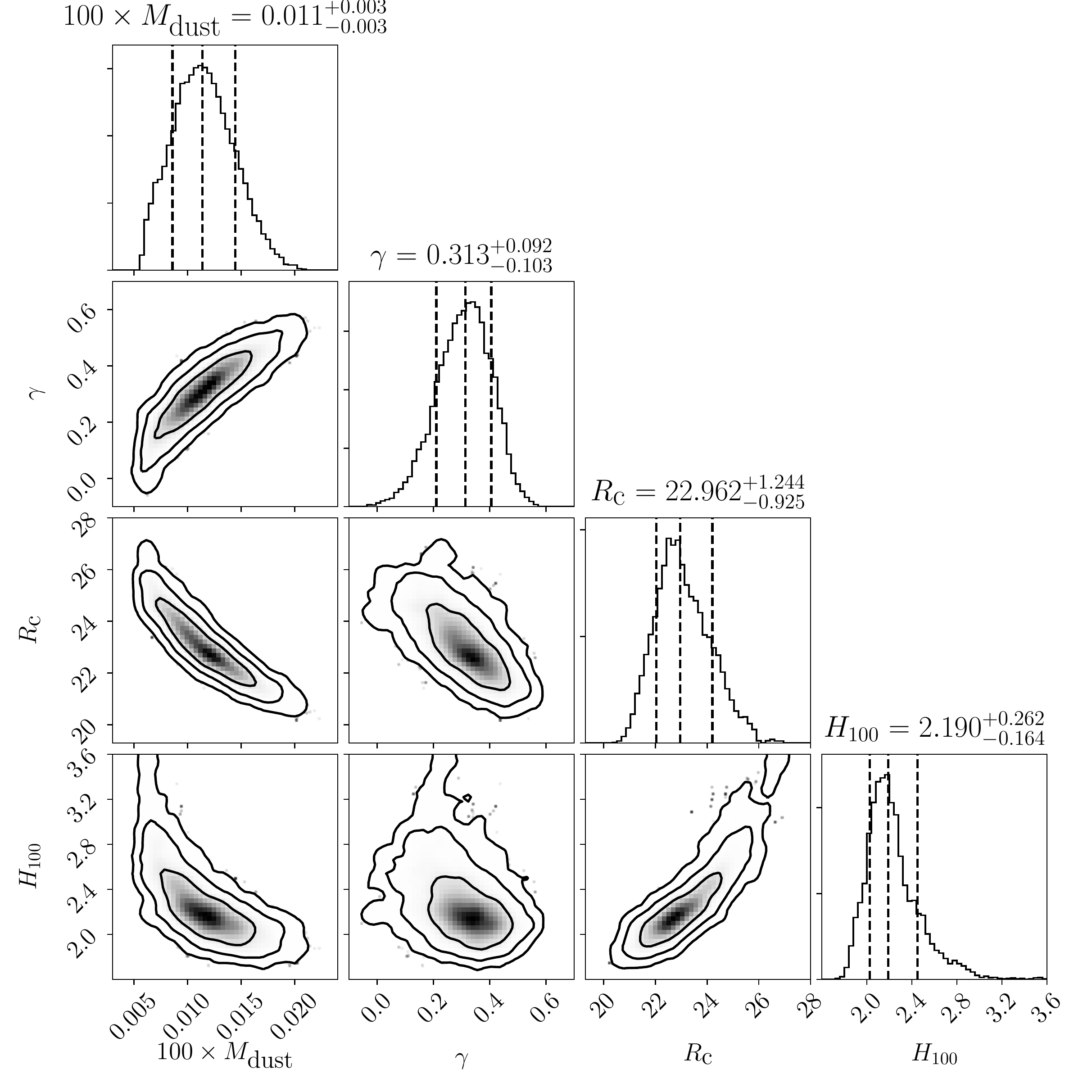}
\caption{Posterior distributions of each dust disc parameter, including
  their marginalized distributions for EX~Lup. The vertical dashed
  lines represent the 16$^{\rm{th}}$, 50$^{\rm{th}}$ and 84$^{\rm{th}}$
  percentiles. Contours correspond to the 68\%, 95\%, and
  99.7\% confidence regions.
}\label{fig-mcmc}
\end{figure*}

The best-fitting parameters and their uncertainties were obtained
after running 1000 iterations ($\sim$~10 times the autocorrelation
time) with 500 walkers. The posterior distributions of $M_{\rm dust}$,
$\gamma$, $R_{\rm c}$, and $H_{100}$ are presented in
Fig.~\ref{fig-mcmc}.

\subsection{Comparison with Axisymmetric CO Gas Disk Model}\label{co-models}

In this section we compare the EX~Lup line observations to an
axisymmetric CO gas disk model, with the purpose of highlighting the
non-symmetric $^{12}$CO(2--1) line emission as well as reproducing the
$^{13}$CO(2--1) and C$^{18}$O(2--1) emission. For this purpose, we use
the three-dimensional Monte Carlo radiative transfer code {\sc mcfost}
\citep{pinte2006,pinte2009} to produce model images of
$^{12}$CO(2--1), $^{13}$CO(2--1), and C$^{18}$O(2--1) for a 
disk model that yields reasonable agreement with the observed
$^{13}$CO and C$^{18}$O emission.  We assumed that the low-$J$ CO
level populations are in LTE at a temperature of $T_{\rm gas}=T_{\rm
  dust}$ in order to produce line emission data cubes. The abundance
of $^{12}$CO relative to H$_2$ was set to 10$^{-4}$ and assumed
constant throughout the disk. Freeze-out of CO molecules onto dust
grains was assumed to occur at temperatures below 20~K with a
depletion factor of 1000.

\reply{We used a genetic algorithm
  \citep[e.g.][]{mathews2013,pinte2018} to fit the $^{13}$CO and
  C$^{18}$O moment 0 maps simultaneously. The CO disk model does not
  attempt to fit the non-axisymmetric $^{12}$CO(2--1) line emission,
  which is beyond the scope of this paper. Each generation was
  composed of 100 models and the genetic algorithm was run for 50
  generations. The following parameters were allowed to vary during
  the fitting process: $M_{\rm gas}$, $\gamma$, $R_{\rm in}$,
  H$_{100}$ as well as the disk's flaring exponent.  Initially $R_{\rm
    c}$ and the disk's outer radius $R_{\rm out}$ were also part of
  the variable parameters, but we found that better solutions were
  obtained if we restricted $R_{\rm c}$ to 75~au and the outer radius
  to 150~au. To determine these values, a range of $R_{\rm c}$ between
  50 and 90~au and $R_{\rm out}$ between 100 and 200~au were explored
  by eye, in steps of 5~au and of 10~au, respectively. Furthermore,
  because the algorithm does not fully sample the parameter space, the
  solution found by the genetic algorithm may not be unique and
  therefore one must be careful when interpreting the model parameters
  presented in Table~\ref{tablemodel1}.

The CO disk model is different from the dust disk model presented in
Section~\ref{models}: both the dust and CO disk models assume the same
total dust mass from Section~\ref{models} but the radial distribution
of dust in the CO disk model is assumed to be more radially extended,
as described by the characteristic radius parameter $R_{\rm c}$ for
the gas model. Differences in the radial distribution millimeter-sized
dust grains and CO gas have been reported in protoplanetary disk
systems \citep{deGregorio2013,pinte2016}, and can be explained by
faster grain growth in the central regions and/or inward radial
migration of larger dust grains \citep{barriere2005,pinte2014}.}

\begin{table*}
  \caption{EX~Lup CO Disk Model}
  \label{tablemodel1}
  \begin{center}
    \leavevmode
    \begin{tabular}{lcc} \hline \hline              
   Physical Parameter  &   &  Reference   \\
 \hline    \hline          
  Stellar Properties &   &  \\
\hline   
Primary Star Mass: $M_\star$ (M$_{\odot}$)  &  0.5  & \citet{frasca2017}  \\
Effective Temperature: $T_\star$ (K) &  3859   &  \citet{frasca2017} \\
Stellar Radius: $R_\star$ (R$_{\odot}$) &  0.8  &  \citet{frasca2017} \\
Distance: $d$ (pc) &  155 &   \citet{lombardi2008}  \\
 \hline 
  Disk Structure\tablenotemark{a}   &  Gas & Dust \\
 \hline 
Disk Dust Mass: $M_{\rm d}$  (M$_{\odot}$)  &   $1.1\times 10^{-4}$     &  $1.1^{+0.3}_{-0.3}\times 10^{-4}$ \\
Disk Total Mass: $M_{\rm D}$ (M$_{\odot}$) &  $ 5.1\times10^{-4}$    & 100 $\times$ $M_{\rm d}$ \\
Inner Rim Radius: $R_{\rm in}$ (au)  &  0.27   & 0.05  \\
Characteristic Radius: $R_{\rm {c}}$ (au) &  75   & $22.9^{+1.2}_{-0.9}$  \\
Outer Radius: $R_{\rm{ out}}$ (au) &  150   & -  \\
Characteristic Height at 100 au: $H_{\rm{c}}$ (au) &  7.6  &   $2.2^{+0.3}_{-0.2}$ \\
Surface Density Exponent: $\gamma$ &   0.32 & $0.31^{+0.09}_{-0.10}$  \\ 
Flaring Exponent: $\psi$ &   1.0  & 1.1  \\ 
Inclination Angle: $i$ ($^{\circ}$)  &  38   &  32 \\
Position angle: PA ($^{\circ}$)  &  78   &  65 \\
Systemic Velocity: $\rm{V}_{\rm{LSR}}$ (km~s$^{-1}$) & +4.1    &   \\
 \hline  
    \end{tabular}
    \tablenotetext{a}{Uncertainties are only given for parameters fit in the MCMC (see Section~\ref{model}).}
  \end{center}
\end{table*}

\begin{figure*}
\includegraphics[width=1.\textwidth]{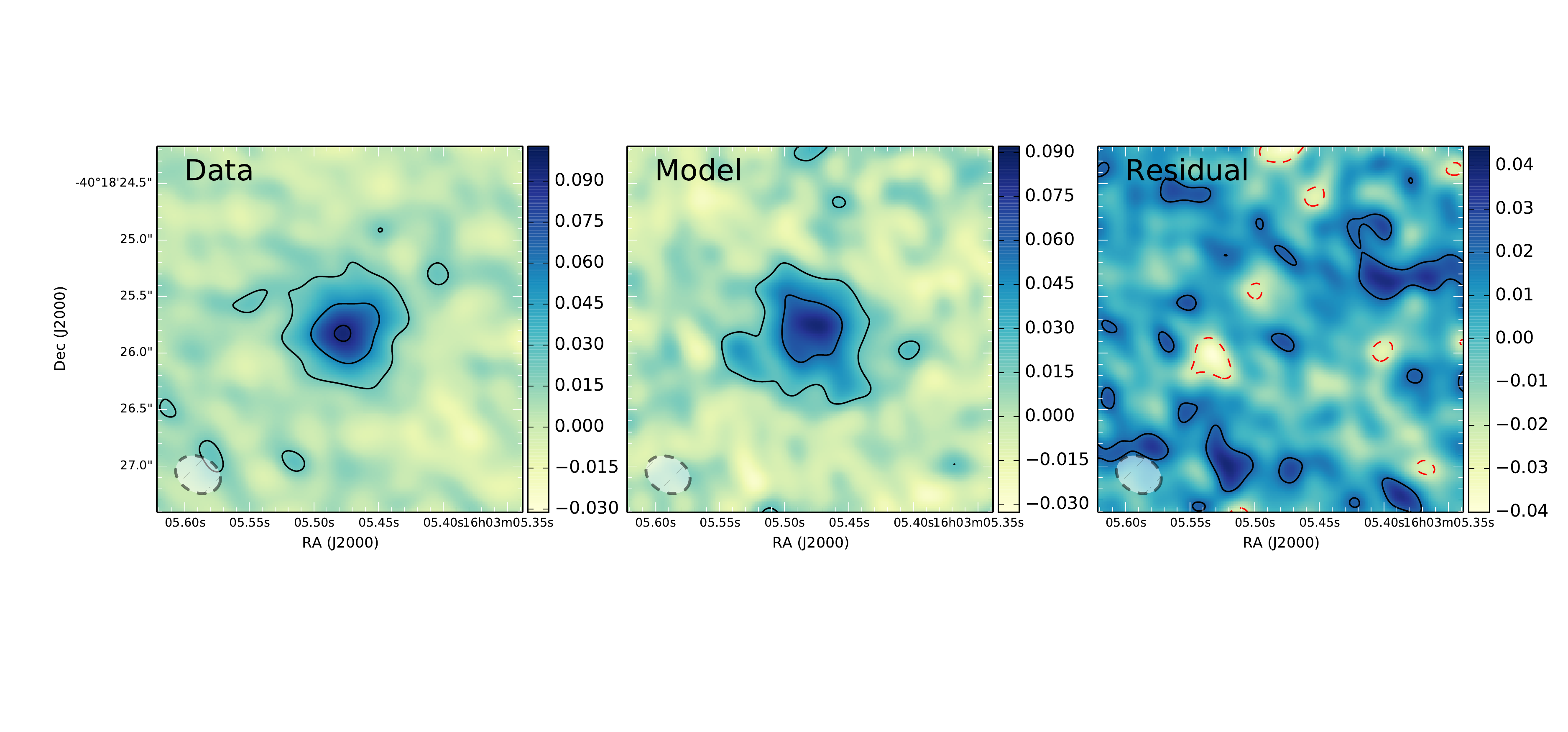}
\caption{{\it Left to right:} $^{13}$CO(2--1) moment 0 images for
  data, model, and residual, respectively. Contour levels are same as
  Figure~\ref{fig-1}.}\label{fig-moment0-13co}
\end{figure*}
\begin{figure*}
\includegraphics[width=1.\textwidth]{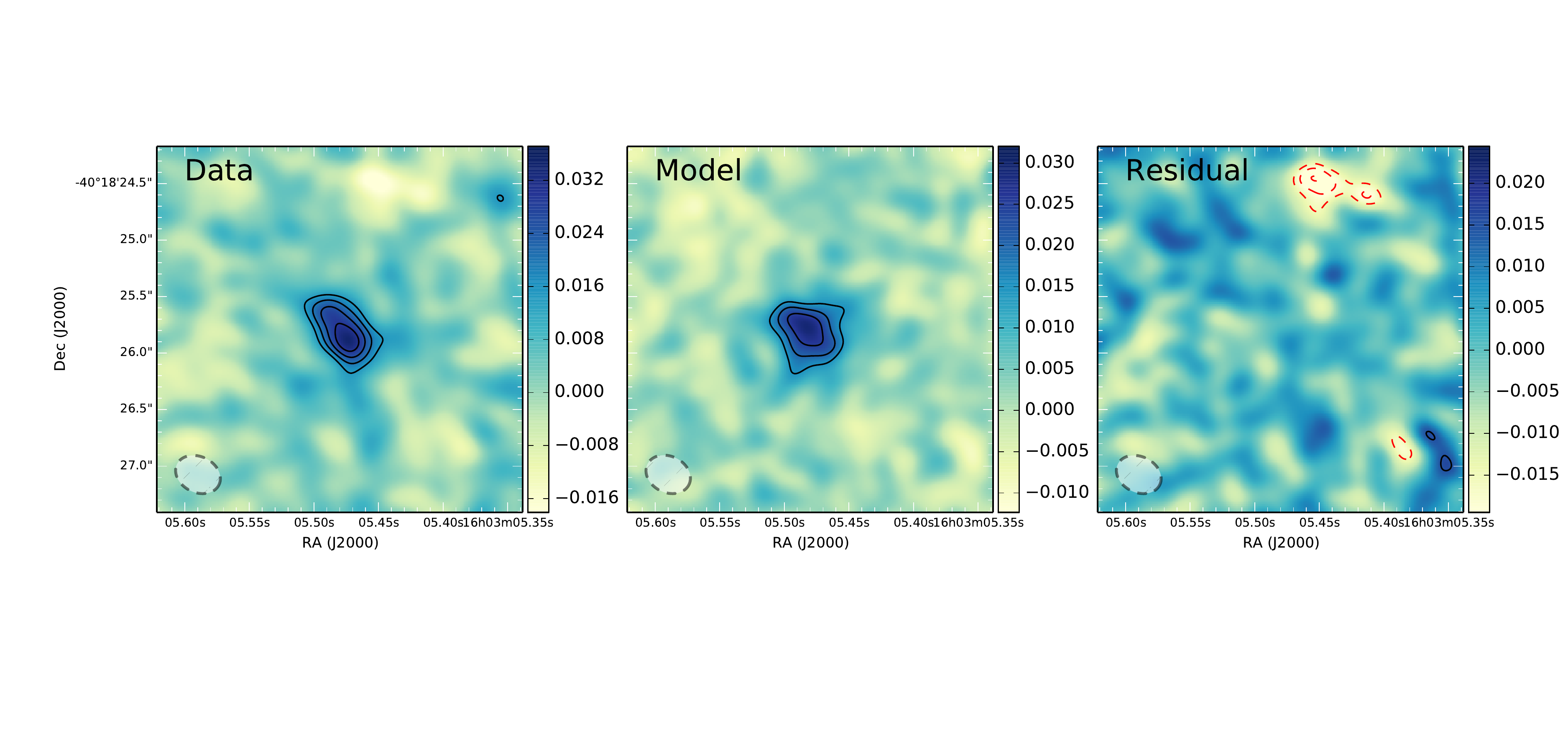}
\caption{{\it Left to right:} C$^{18}$O(2--1) moment 0 images for
   data, model, and residual, respectively. Contour levels are same as
  Figure~\ref{fig-1}.}\label{fig-moment0-c18o}
\end{figure*}

The model images were used to create synthetic visibilities using CASA
task {\sc simobserve}, using the same integration time, spectral setup
and antenna configuration used for the observations as well as
injecting the appropriate amount of thermal noise. The resulting model
visibilities were then imaged using the same {\sc clean} parameters
used for the real data. Figures~\ref{fig-moment0-13co} and
~\ref{fig-moment0-c18o} show the resulting Moment 0 images for the
data, model, and residuals for $^{13}$CO(2--1) and C$^{18}$O(2--1)
respectively.  The model reproduces well the observed $^{13}$CO(2--1)
and C$^{18}$O(2--1) integrated intensity images. \reply{We find that a
  vertical optical depth of 1 is reached at a radius of 60~au for
  $^{13}$CO(2--1) and 15~au for C$^{18}$O(2--1).  $^{12}$CO(2--1) is
  very optically thick throughout the disk, and has an optical depth
  of 1.5 even at the outer radius of 150~au.}

%

\section{Discussion}\label{discussion}
\subsection{Dust Disk}

The ALMA observations of EX~Lup resolve the dust disk for the first
time. The total dust mass derived from the radiative transfer model fitting process is
1.0$\times10^{-4}$\,M$_\odot$ (33.3\,M$_\earth$). Our model predicts,
however, that the disk becomes optically thick at $r<10$~au, so the
total dust mass could be higher by a factor of 2--5. Deriving the dust mass using the
standard assumption of optically thin emission, a dust temperature of
20~K, and the opacities of \citet{beckwith1990}, underestimates the
dust mass by a factor of at least 3 compared to our model. The model predicts an
0.87~mm flux of $\sim$50~mJy, consistent with the flux measured with LABOCA
\citep{juhasz2012}. The 1.3~mm spectral index measured in-band and also between 0.87
and 1.3~mm are both close to the value of 2.3 measured toward FU~Ori itself
\citep{liu2017}. The low spectral index of FU~Ori can be explained by the presence of
a compact optically thick inner region surrounded by a more extended optically thin
disk \citep{liu2017}. A similar spectral behavior has been reported in the bright FUor
system V883~Ori \citep{cieza2016}, which shows an axisymmetric optically thick inner
disk within 42~au surrounded by an optically thin halo extending out to 125~au. 
The modeling procedure applied to EX~Lup is similar to the one we performed for other
FUor/EXor sources \citep{cieza2017}, allowing direct comparison. EX~Lup has an
intermediate dust mass compared to other FUor and EXor objects
\citep{liu2016,cieza2017}. The $\sim 23$~au characteristic radius we measure is
consistent with the compact radii measured toward other FUor/EXor
sources. The $\gamma$ value of 0.34 is lower than measured towards most FUor objects
observed by \citet{cieza2017} and closer to the values seen around the FUor/EXor
object V1647~Ori and toward T~Tauri stars \citep{andrews2010}. We find the disk around
EX~Lup follows the trend that EXor stars are surrounded by compact dust disks which
become optically thick in their inner regions \citep{cieza2016,liu2017,cieza2017},
which could be the reservoir feeding the accretion. 

We find no evidence of strong asymmetries in the dust continuum, which
have been suggested to occur as a direct consequence of gravitational
instability \citep[e.g.][]{zhu2012}. \reply{Comparison with an axisymmetric
model suggest the presence of structure not accounted at the 5~sigma
level.} Assuming a gas-to-dust ratio of 100 the total disk mass is
0.011\,M$_\odot$, close to the 0.025\,M$_\odot$ derived with SED
fitting by \citet{sipos2009}. Given that gravitational instability and
disk fragmentation require $M_{\rm D}/M_\star > 0.1$
\citep[e.g.][]{armitage2001,vorobyov2015}, the total disk mass derived
from the continuum is consistent with the non-detection of asymmetries
in the dust disk triggered by gravitational instability.

\subsection{CO Gas Disk}

We detect widespread $^{12}$CO(2--1) emission, of which some fraction
can be attributed to a Keplerian gas disk. \citet{williams2014}
developed a grid of protoplanetary disk models that can be used to
derive disk masses based on the $^{13}$CO(2--1) to C$^{18}$O(2--1)
integrated line ratios (the models take into account
photodissociation, CO freeze-out, and basic CO chemistry). Comparing
the EX~Lup $^{13}$CO/C$^{18}$O line ratios to those from the models of
\citet{williams2014}, we obtain a total disk mass of
5.4$\times10^{-4}$\,M$_\odot$. \reply{This result is remarkably
  similar to the 5.1$\times10^{-4}$\,M$_\odot$ we derived by modeling
  of the $^{13}$CO and C$^{18}$O moment 0 images in
  Section~\ref{co-models}. This is an order of magnitude lower than
  1-3$\times10^{-3}$\,M$_\odot$ total disk mass derived by
  \citet{Sicilia2015} from revised SED fitting, but close to the value
  of 2.3$\times10^{-4}$\,M$_\odot$ derived by \citet{kospal2016b}
  based upon the single-dish detection of the $^{13}$CO(3--2) line. We
  note that since in our model the $^{13}$CO and C$^{18}$O become
  optically thick at radii of 60 and 15~au respectively, the resulting
  gas mass estimate may only represent a lower limit to the total gas
  mass.}  Combined with the dust modeling results, our gas mass
estimates imply a gas-to-dust ratio of 4.5, which is similar to the
value reported by \citet{kospal2016b}.  Gas to dust ratios lower than
the typical ISM value of 100 using the \citet{williams2014} seem to be
common around disks in Lupus and Chameleon
\citep{ansdell2016,villenave2018}. \reply{Depletion of CO and other
  molecular species by up to an order of magnitude during accretion
  outbursts in EX~Lup have been detected through mid-infrared
  spectroscopy \citet{banzatti2015}, and could possibly be related to
  the low gas-to-dust ratios we derived based on the $^{13}$CO and
  C$^{18}$O lines}.

\begin{figure*}
\centering\includegraphics[width=1.\textwidth]{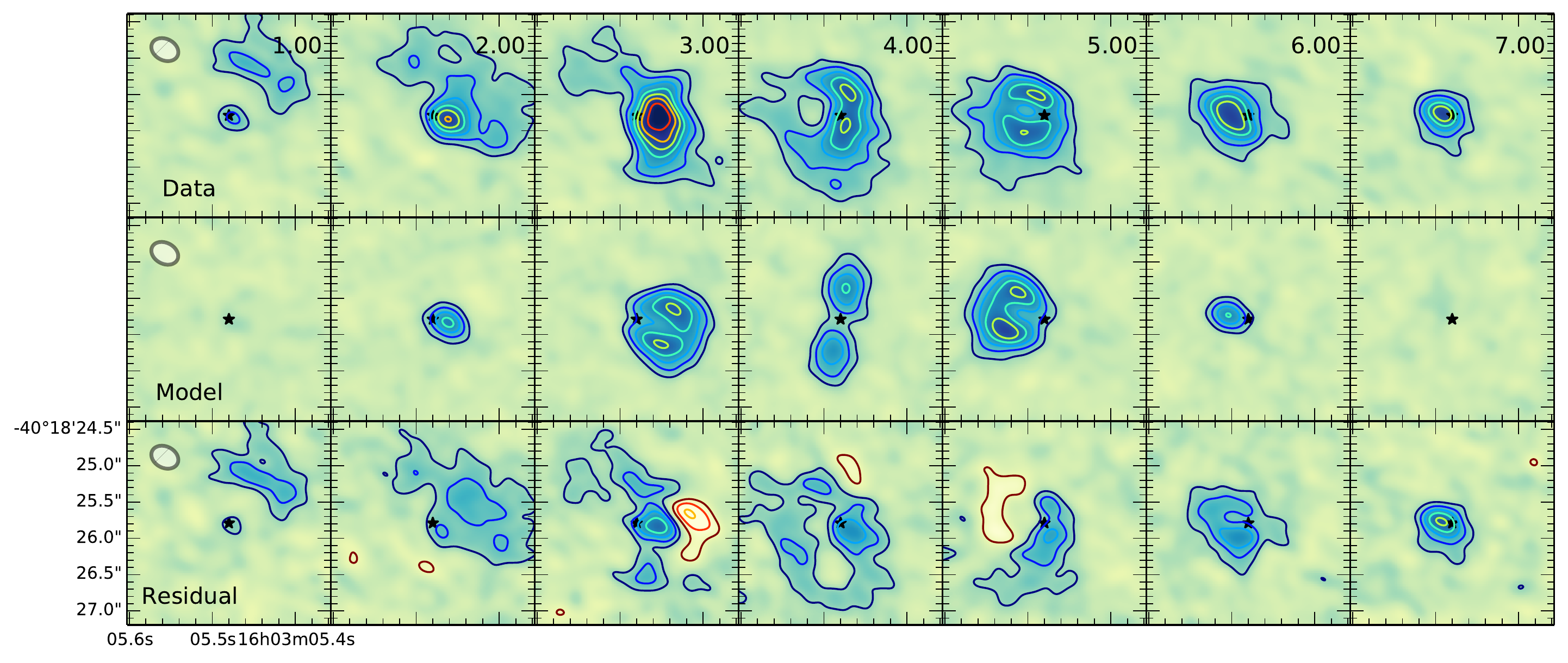}
\caption{Channel maps of the data, axisymmetric model and residuals
  for the $^{12}$CO emission. Contour levels and intensity scale are
  identical as for Figure~\ref{fig2_12co}, but including negative
  contours to highlight negative residuals. The star symbol in the
  center of each panel represents the stellar position. \reply{The
    beam is shown in the top left corner of the leftmost
    panels.}}\label{fig_12co_mod_res}
\end{figure*}

\subsection{Widespread $^{12}$CO(2--1) Gas Emission}

We detect $^{12}$CO(2--1) emission up to the full extent of the
MRS ($\sim 3.2\arcsec$), suggesting there could
be larger-scale emission filtered out by the interferometer.
Unfortunately, there are no $^{12}$CO(2--1) single-dish
observations available which could have been used to estimate the amount of
resolved-out emission. 

High accretion rates in EXor systems must be accompanied by high
mass-loss rates \citep{hartmann2008,hartmann2016}. Prominent
collimated outflows, as seen around FUor objects
\citep{ruiz2017,zurlo2017,kospal2017b,feher2017,principe2018}, are not
evident in the EX~Lup data. However, we identify an arc-like feature
in $^{12}$CO blue-shifted emission, 100--200~au north-west of the star
(see Fig.~\ref{fig2_12co} and Fig.~\ref{fig-12co-mom-vels}). This
emission is detected between +0.6 and +1.8\,km\,s$^{-1}$, i.e.,
displaced to the blue of the systemic velocity by at least
$\sim$~2.0\,km\,s$^{-1}$. This gas and its corresponding velocity
gradient are both in a direction perpendicular to the disk's major
axis and we conclude that this blue emission corresponds to a
molecular outflow, possibly interacting with ambient
material. \reply{We expect this CO emission more likely to pertain to
  swept up molecular material rather than material launched directly
  from the source during the recent burst, because the latter scenario
  would require velocities of 100\,km\,s$^{-1}$ in order for the
  material to have traveled 150~au in 8 years.}

Accretion-related winds have been
detected through optical spectroscopy in EX~Lup during the periods of
enhanced accretion and are believed to originate in the inner $r~<~0.4$~au region of
the disk \citep{kospal2011,goto2011,Sicilia2012}. This probable detection of an
outflow provides evidence that the bipolar outflow phenomenon persists into the EXor
phase, and that accretion/outflow is still clearing molecular gas from the the areas
above and below the disk plane. EXor/FUor sources are therefore likely to be valuable
for studying episodic infall/accretion/outflows near the end stages of the outflow phenomenon. These larger-scale outflow phenomena are still relevant to the evolution
of the disk, as they regulate the accretion process via removal of angular momentum.

We do not detect a clear red-shifted counterpart to the blue-shifted outflow. There is
red-shifted emission at high velocities close to the star in excess of our Keplerian
disk model which is not seen on the blue side. This could be either material
outflowing or infalling towards the star, although we believe it is most likely the red-shifted counterpart of the molecular outflow.

This interpretation places the disk's far (near) side to the north-west
(south-east) with the exposed surface rotating clockwise as seen by the observer.
This is consistent with the observed disk kinematics, best evidenced in the C$^{18}$O
first moment map (Figure~\ref{fig-mom1}). The proposed geometry for the disk and
molecular outflow is shown in Figure~\ref{schematics}. This geometry is consistent
with the inner disk configuration proposed by \citet{Sicilia2012}.

\begin{figure}
\centering\includegraphics[width=0.5\textwidth]{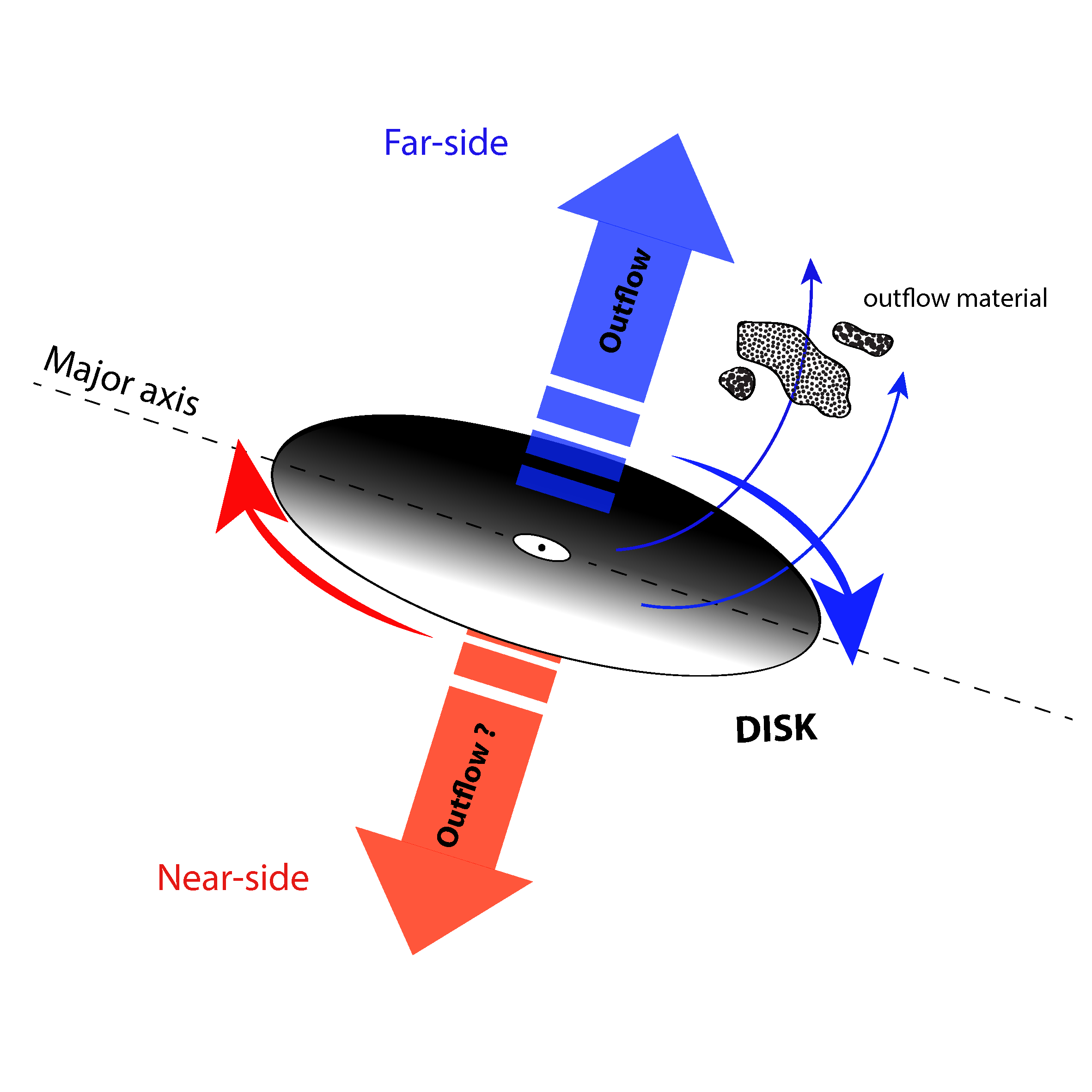}
\caption{Schematic diagram of the EX~Lup system.}
\end{figure}\label{schematics}
      
Molecular outflows and complex circumstellar structures have
been observed in FUor sources and in the borderline FUor/EXor object
V1647~Ori, but not in EXors: this has led to the suggestion that FUor
objects may be precursors to EXors, with EXors corresponding to a
slightly more evolved stage \citep{hartmann2008,cieza2017}. EX~Lup
itself seems to have a moderately complex circumstellar environment,
with evidence of recent outflowing material still impacting the
tenuous remnants of the ambient cloud and extended material near
systemic velocities possibly still spiraling into the
system. Compared with the rich, active environments of FUor sources
and the more quiescent environments of the few surveyed EXor objects,
EX~Lup itself may be a system in a transitional phase between
classical FUors and EXors. Ironically, EX~Lup
and FU~Ori seem more similar to each other in terms of disk size and dust mass than
FU~Ori is to other FUor sources or EX~Lup is to other EXors
\citep{hales2015,liu2016,liu2017,cieza2017}.

Fainter extended material at low velocity is detected on the red-shifted side
of the disk (Fig.~\ref{fig-pv-co}). This extended material has a
radial velocity profile $\propto r^{-2.9}$, much steeper than the
nearly $r^{-0.5}$ Keplerian dependence found in the inner
$r~<~100$~au, and also steeper than the $r^{-1}$ gradient
characteristic of flattened, infalling envelopes \citep[e.g][and
references therein]{kospal2017b}. It must be noted that the
red-shifted emission at larger radii is fainter (at the limit of the
MRS), and thus the fit to the centroid positions is less
accurate. This diffuse extended emission is likely associated with a
more tenuous rotating envelope, which has to be low density to be
consistent with the low-reddening and disk-like SED of EX~Lup.

The first moment map also shows a twist around the systemic velocity,
similar to those found in sources where material is still being
transferred from the envelope into the disk
\citep{2017arXiv170802384Y}. This twist corresponds to an
spiral-shaped structure that seems to cross the disk at
velocities between +2.6 and +3.4~km~s$^{-1}$. The two
sides of the spiral cross the disk at center of the blue-side
asymmetry. However, interpretation of structure near the systemic
velocity may be confused by extended cloud emission larger than the MRS,
causing problems during the image reconstruction.

\begin{figure*}
\centering\includegraphics[width=0.9\textwidth]{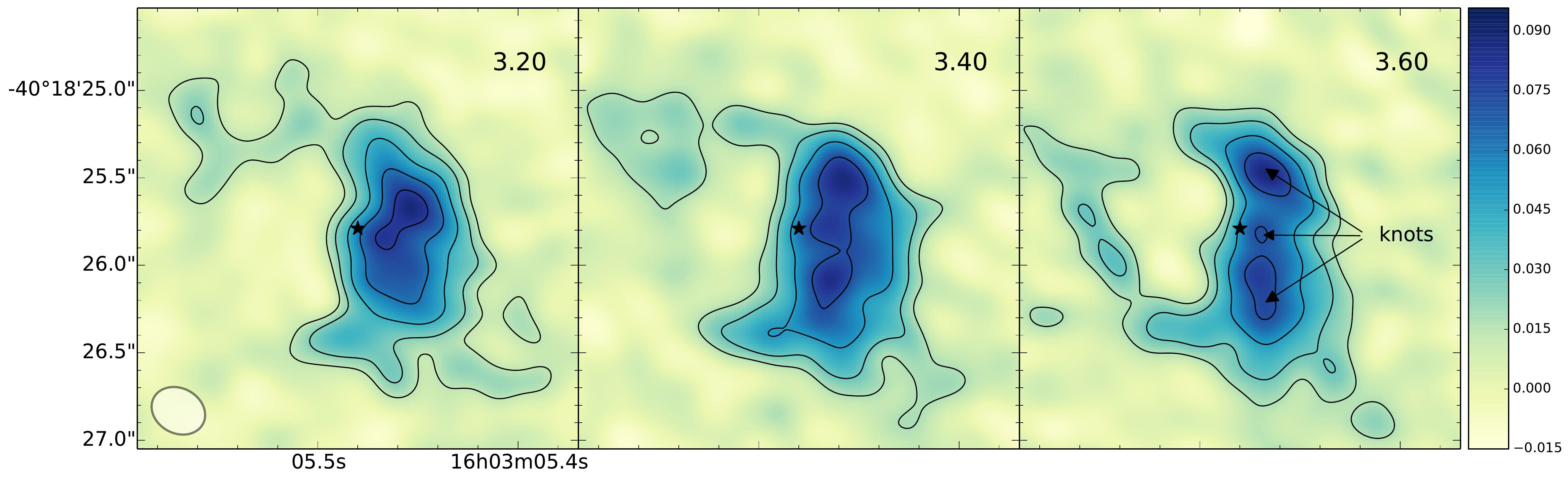}
\caption{Blow-up of the central velocity channels, highlighting the
  clumpy, non-Keplerian $^{12}$CO emission on the blue-shifted
  side. The star symbol in the center of each panel represents the
  stellar position, coinciding with the continuum peak.}\label{fig2_12co_zoom}
\end{figure*}

\subsection{Nature of the Asymmetry and Connection with the EXor Outburst}\label{asymmetry}

The $^{12}$CO(2--1) channel maps reveal clear deviations from
Keplerian kinematics when compared to an axisymmetric disk model
(Fig.~\ref{fig_12co_mod_res}). The ``butterfly'' pattern that is
clearly seen on the red-shifted side of the disk should be symmetric
on the blue-shifted side with respect to the systemic velocity. If the
systemic velocity is assumed to be around +4.1~km~s$^{-1}$, then for an
axisymmetric disk the channel maps at +3.0 and +5.0~km~s$^{-1}$ should
be mirror images of each other. Instead they are very different, with
the +3.0~km~s$^{-1}$ map being more compact than its +5.0~km~s$^{-1}$
red-side counterpart. The velocity channels at +3.2, +3.4, and
+3.6~km~s$^{-1}$ show an elongated, hourglass shape that seems to
cross the disk in the north--south direction. The hourglass shape does
not cross through the disk's center (i.e. the stellar position and
centroid of the continuum), but at $\sim 0.08\arcsec$ to the
south-west ($\sim$1/3 of the beam). Figure~\ref{fig2_12co_zoom} shows
a blow-up of the channel maps near systemic velocity of the
$^{12}$CO(2--1) image cube created using Briggs weighting. The
slightly higher resolution compared to natural weighting shows that
the hourglass shape may consist of at least three unresolved clumps
connected by filamentary emission.

One possible scenario to explain the morphology seen on the blue side
is that we are observing an asymmetry related to the outflow, the
offset corresponding to the launching radius of the disk wind. The
spiral seen near the systemic velocity seems to ``land'' at this
position, consistent with the asymmetry coinciding with
the landing site of infalling material from the remnant envelope and
possibly to the launching site of a disk outflow. This is similar to
what has been reported by \citet{alves2017} for the Class~I object
BHB07-11. Although more massive (1.7~M$_{\odot}$) and
younger than EX~Lup, ALMA observations of BHB07-11 detect bipolar
outflows launched symmetrically from the disk at 80~au radius
from the star, where larger scale spirals and the disk intersect. In
the case of EX~Lup, the center of the two-lobed structure seen at
+3.6~km~s$^{-1}$ is located at 12--15~au, requiring velocities of
$\sim 10$~km~s$^{-1}$ to escape the system.

Alternatively, twisted iso-velocities similar to the morphology seen
at +3.6~km~s$^{-1}$ are associated with enhanced accretion onto the
central star \citep{casassus2015,rosenfeld2014}. In this scenario the
twisted iso-velocities should be centered at the stellar position:
however in our data the center of the twisted hourglass is shifted by
about 1/3 of a beam from the star. Such a shift in the accretion
center could plausibly be the result of an unseen companion. Many
efforts have been made to identify a possible binary companion around
EX~Lup \citep{ghez1997,bailey1998,kospal2014}. \citet{kospal2014}
detect a radial velocity signal that can be described by a companion
with an $m\, {\rm sin}\, i = 14.7 \pm 0.7$~M$_{\rm Jup}$
(0.018~M$_{\odot}$ assuming an inclination of 50$^{\circ}$ located
0.06~au from the primary. \citet{Sicilia2015} however, show that the
same signal can also be reproduced by a rotating, line-dependent
veiling from an accretion column, without the need of a stellar
companion. In addition, a Jupiter-mass companion at 0.06~au may be too
close-in to explain the shift in the $^{12}$CO emission centroid.

A similar twisted feature has been proposed as a signature of on-going
planet formation by \citet{perez2015b}. As an accreting giant planet
develops its own circumplanetary disk, it distorts the global velocity
field due to its own distinct kinematics and accretion streams. This
circumplanetary disk also forces some of the gas to follow horseshoe
orbits, distorting the Keplerian pattern along its full orbit,
producing a large-scale skewed pattern in the iso-velocities, similar
to the one seen in EX~Lup's systemic velocity channels \citep{Perez}.

Asymmetries in the inner disk of EX~Lup have been reported previously
\citep{goto2011,Sicilia2012}, associated with a hot spot in the inner
0.4~au of the disk. Our observations indicate the gas disk is also
asymmetric at outer radii ($>~100$~au), suggesting a possible link between the outer
disk structure and the accretion from the inner regions into the star. In
this picture, the asymmetry observed in the blue regions
corresponds to the region of the disk where low-density material from the envelope
falls into the disk and piles up until it can be efficiently
transferred inwards. The infall dragging of the magnetic field lines
can trigger the launching of disk winds \citep{zhao2016}, which may
enhance the asymmetry between red-shifted and blue-shifted sides. A
mismatch between infall from the envelope and accretion onto the star
has been suggested as the possible cause for the piling-up of material
in the disk, leading to disk instabilities and enhanced accretion
\citep{bell1994}. Observational evidence of this has
recently been reported in the FUor star V346~Nor by \citet{kospal2017b}
using ALMA observations incorporating Morita
Array short-baseline and single-dish observations.
This combination of data from different
array families allows characterization of the circumstellar environment
on different spatial scales and to constrain
better the envelope infall rates. Our ALMA 12~meter array CO observations of EX~Lup
suggest the presence of extended emission, but observations
targeting larger-scale emission are needed to quantify the balance
between envelope mass infall and how material from the asymmetric outer
disk is transferred into the inner disk and ultimately onto the star.

The clumpy substructure detected in CO inside the region of the
asymmetry could represent the clumps and spiral filaments predicted by
models of fragmentation in gravitationally unstable disks
\citep{vorobyov2015}. The models predict these features should be
detectable in the gas and that the interaction between disk and clump
fragments could be responsible for the non-periodic accretion
outbursts. The current total disk mass estimates suggest, however,
that the disk is not massive enough for gravitational instabilities to
occur. Moreover the dust disk appears less asymmetric than might be
expected. Detailed hydrodynamic modeling and higher-resolution
observations of the EX~Lup disk are required to quantify whether the
observed structure could be explained by fragmentation driven by
gravitational instability.

\section{Conclusions}\label{conclusion}

We present the first millimeter-wave resolved observations of the
prototype EXor outbursting system EX~Lup using ALMA. We detect and
resolve emission from EX~Lup's disk in both continuum and the CO
isotopologues $^{12}$CO(2--1), $^{13}$CO(2--1), and
C$^{18}$O(2--1). Our main findings are:

\begin{itemize}

\item The continuum disk is compact, with a characteristic radius of
  23~au and a total dust mass of 1.0$\times10^{-4}$~M$_\odot$
  (33\,M$_\earth$).  Our modeling shows that the 1.3~mm continuum is
  optically thin beyond $\sim$~10~au but the data shows no asymmetries
  or sub-structures \reply{above 5-sigma level}.
\item The $^{12}$CO disk is highly asymmetric, with the blue side
  deviating significantly from Keplerian rotation. On the other hand,
  $^{13}$CO and C$^{18}$O trace disk rotation and are used to estimate
  a disk gas mass of 5.1$\times10^{-4}$. The red side of the disk
  follows Keplerian rotation. Assuming a stellar \reply{mass of 0.5~M$_\odot$
  we derive an inclination angle of 38$^{\circ}~\pm~4^{\circ}$}.
\item The overall $^{12}$CO emission shows various components
  additional to the disk, such as extended emission at systemic
  velocities and a blue-shifted molecular outflow at
  100--200~au. The detection of a molecular outflow around EX~Lup
  provides evidence that episodic accretion occurs throughout the early stages of star
  formation and the amount of activity declines gradually with age of the YSOs;
  therefore, the distinction between FUors and EXors is one of degree, not of kind.
\end{itemize}

\noindent The bright asymmetry on the blue-shifted side shows a clumpy, twisted
structure which is likely associated to the launching site of the
molecular outflow. However ruling out other possible explanations such
disk instabilities, disk-companion interactions, or optical depth
effects requires further modeling. Follow-up observations targeting
larger-scale emission are crucial to understanding the mass transport
balance from the remnant envelope onto the disk.

\section*{Acknowledgments}

\reply{We thank the anonymous referee for their insightful review. We thank E. Fomalont and J. Williams for useful discussions and suggestions during the analysis of the data.} This paper makes use of the following ALMA data:
{\sc ADS/JAO.ALMA.2015.1.00200.S}. ALMA is a partnership of ESO
(representing its member states), NSF (USA) and NINS (Japan), together
with NRC (Canada) and NSC and ASIAA (Taiwan), in cooperation with the
Republic of Chile. The Joint ALMA Observatory is operated by ESO,
AUI/NRAO and NAOJ. The National Radio Astronomy Observatory is a
facility of the National Science Foundation operated under cooperative
agreement by Associated Universities, Inc. This research made use of
{\sc astropy}, a community-developed core {\sc python} package for astronomy
\citep{2013A&A...558A..33A}. This research used the facilities of the Canadian Astronomy Data Centre operated by the National Research Council of Canada with the support of the Canadian Space Agency. S.P. and L.C. acknowledge support from the
Millennium Science Initiative (Chilean Ministry of Economy) through
grant Nucleus RC130007. S.P. acknowledges CONICYT-Gemini grant
32130007. L.C. acknowledges support from CONICYT-FONDECYT grant 1171246. This paper used the Brelka cluster, financed by Fondequip
project EQM140101. C.P. acknowledges funding from the Australian Research Council (ARC) under the Future Fellowship number FT170100040. This work was undertaken with the SOLA (Soul of Lupus with ALMA) collaboration centered at the Joint ALMA Observatory. 

\vspace{5mm}

\facilities{Atacama Large Millimeter/Submillimeter Array}


\software{Common Astronomy Software Applications \citep{2007ASPC..376..127M}, {\sc radmc-3d}, \citep{Dullemond2012}, MCFOST, \citep{pinte2006,pinte2009}, Astropy \citep{2013A&A...558A..33A}}


\appendix
\section{Channel Maps}\label{co_maps}

\begin{center}
\begin{figure}[b]
\includegraphics[width=1.0\textwidth]{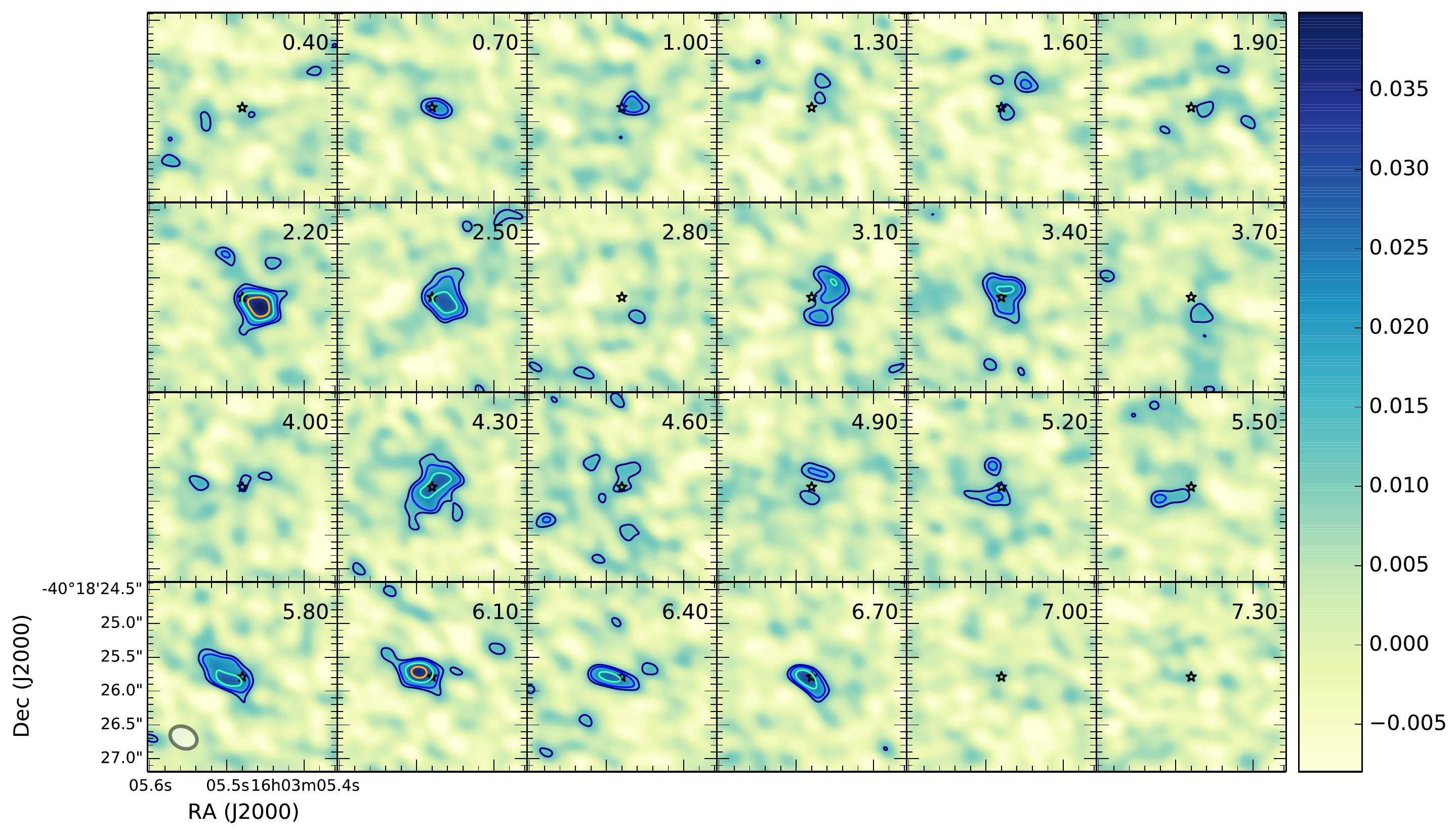}
\caption{$^{13}$CO(2--1) channel maps towards EX~Lup. The velocity of the
  channels is shown in the Local Standard of Rest (LSR) frame,
  centered at the rest frequency of $^{13}$CO(2--1). The data has
  been binned to a velocity resolution of 0.3~km~s$^{-1}$. All maps
  share the same linear color scale. Contour levels are 0.012, 0.016, 0.024, 0.032,
  and 0.04~Jy\,beam$^{-1}$.}\label{fig2_13co}
\end{figure}
\end{center}

\begin{center}
\begin{figure}
\includegraphics[width=1.0\textwidth]{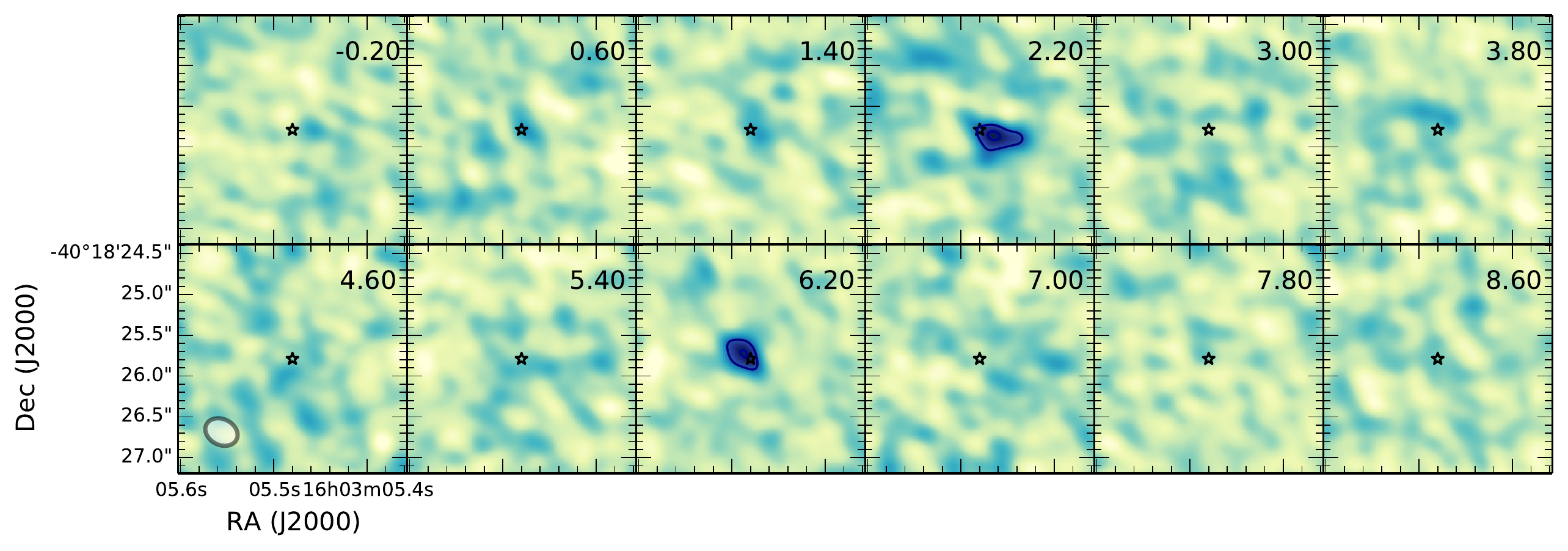}
\caption{C$^{18}$O(2--1) channel maps towards EX~Lup. The velocity of the
  channels is shown in the Local Standard of Rest (LSR) frame,
  centered at the rest frequency of C$^{18}$O(2--1). The data has
  been binned to a velocity resolution of 0.8~km~s$^{-1}$. All maps
  share the same linear color scale. Contour levels are 0.055, 0.010, and
  0.015~Jy\,beam$^{-1}$.}\label{fig2_c18o}
\end{figure}
\end{center}

\begin{figure}
\centering\includegraphics[angle=0,scale=0.25]{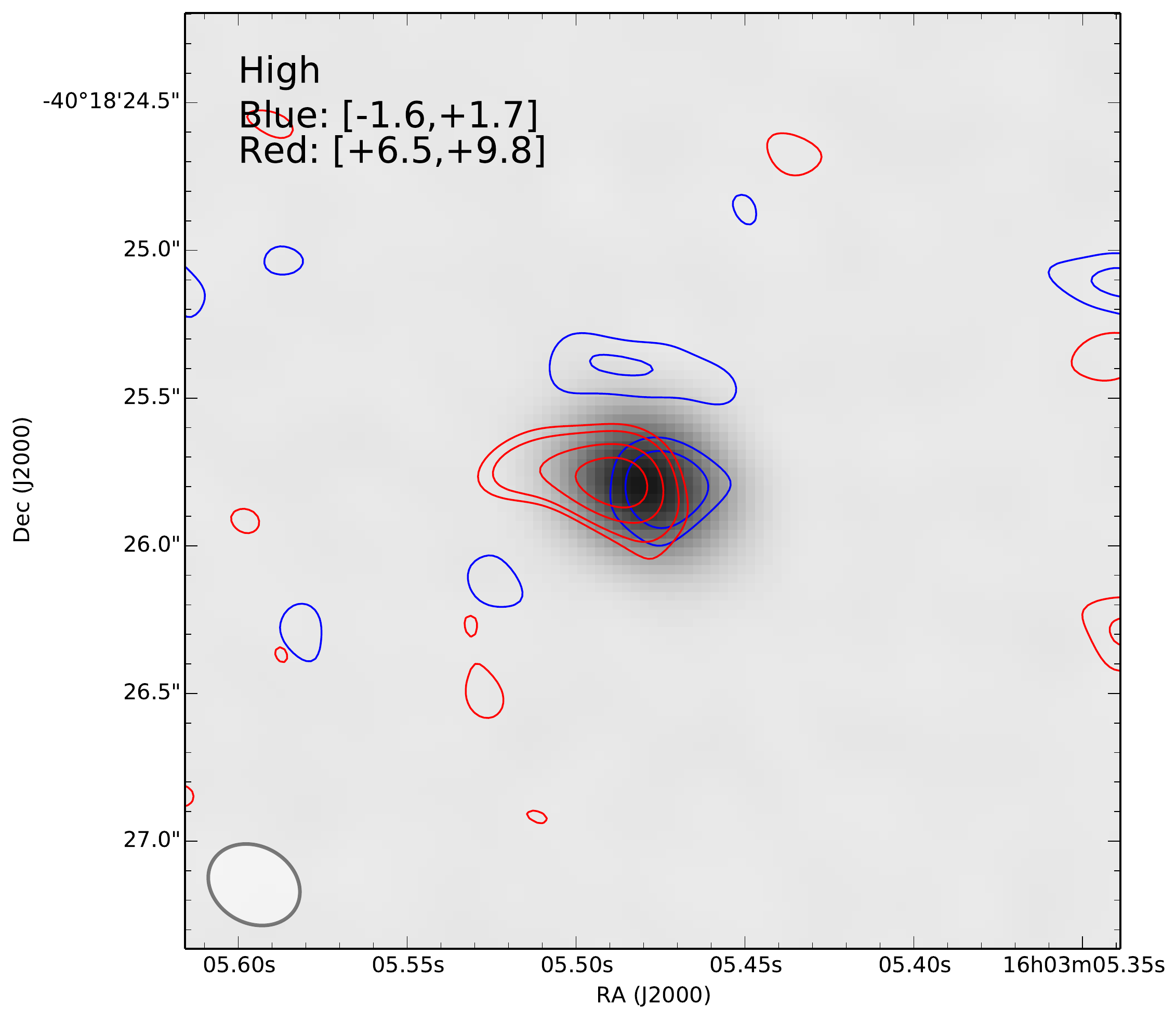}
\centering\includegraphics[angle=0,scale=0.25]{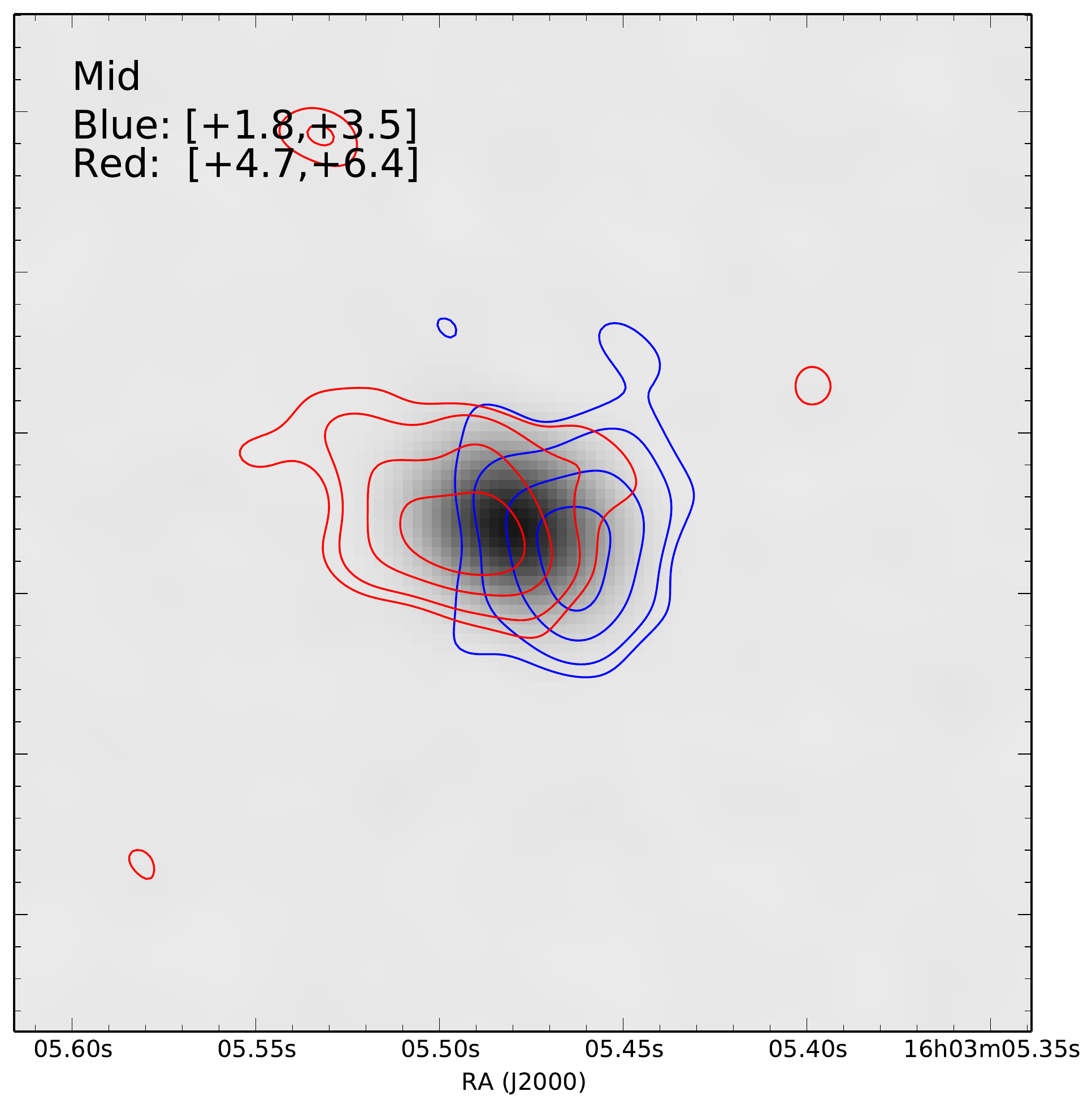}
\centering\includegraphics[angle=0,scale=0.25]{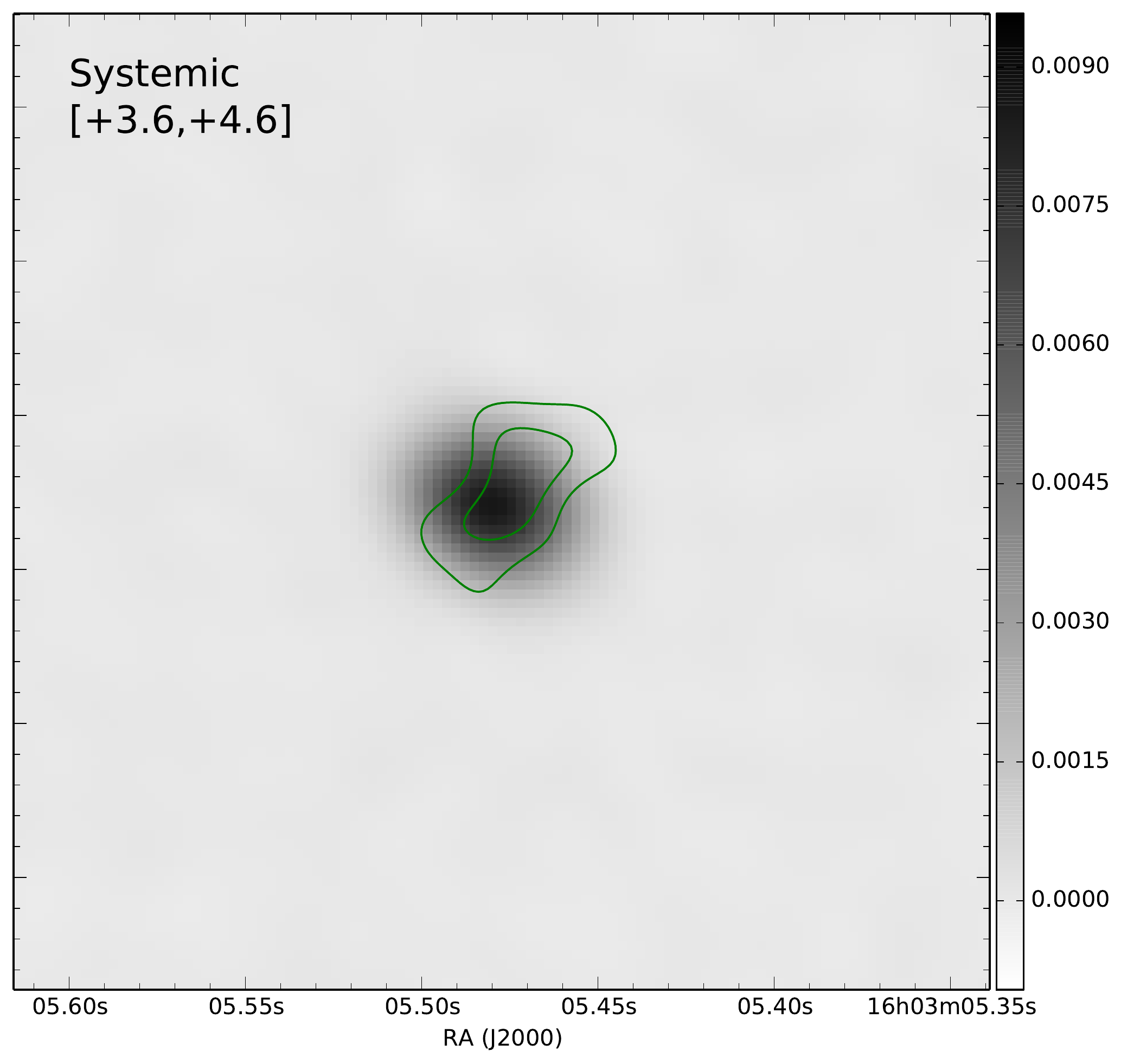}
\centering\includegraphics[angle=0,scale=0.25]{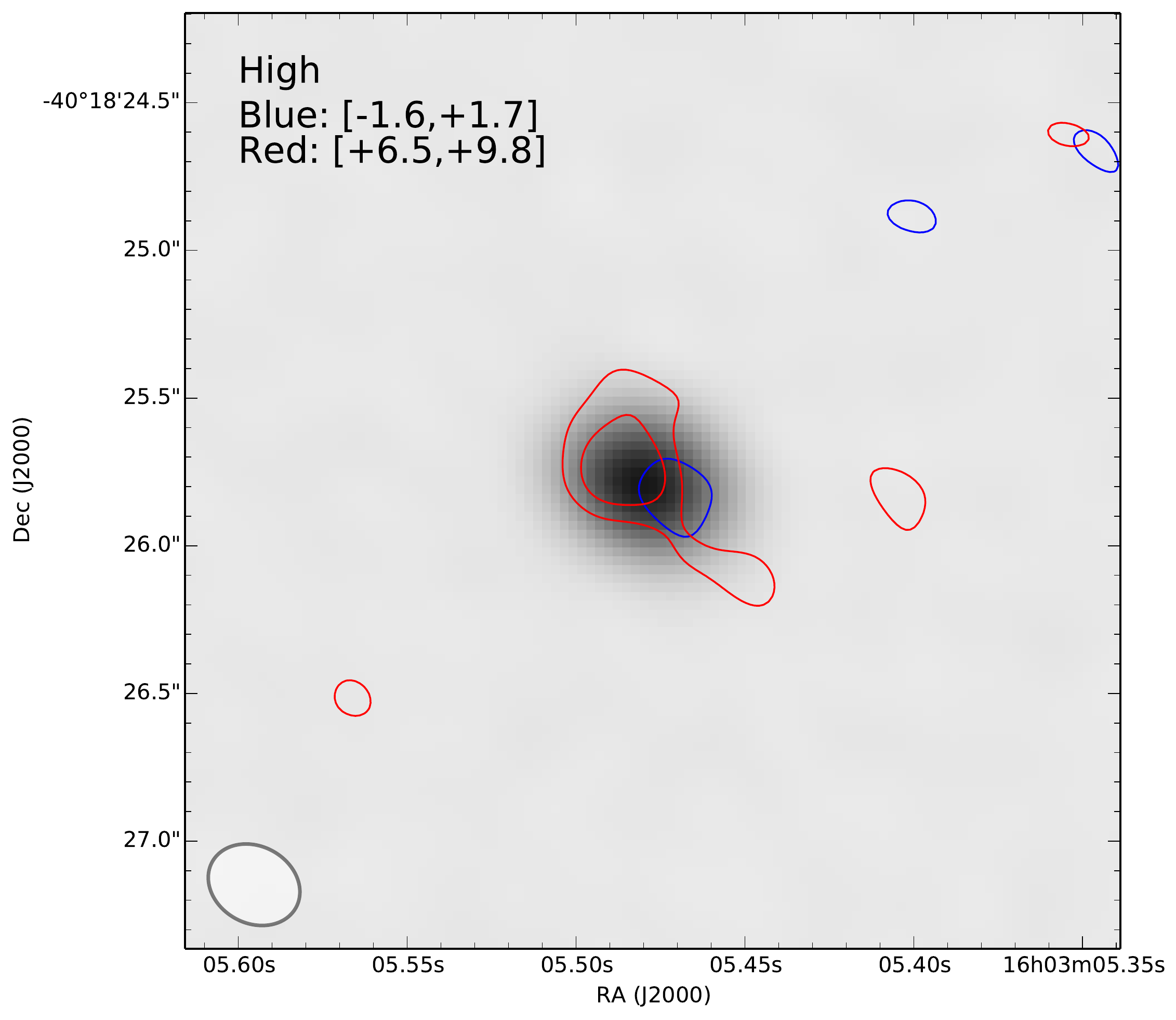}
\centering\includegraphics[angle=0,scale=0.25]{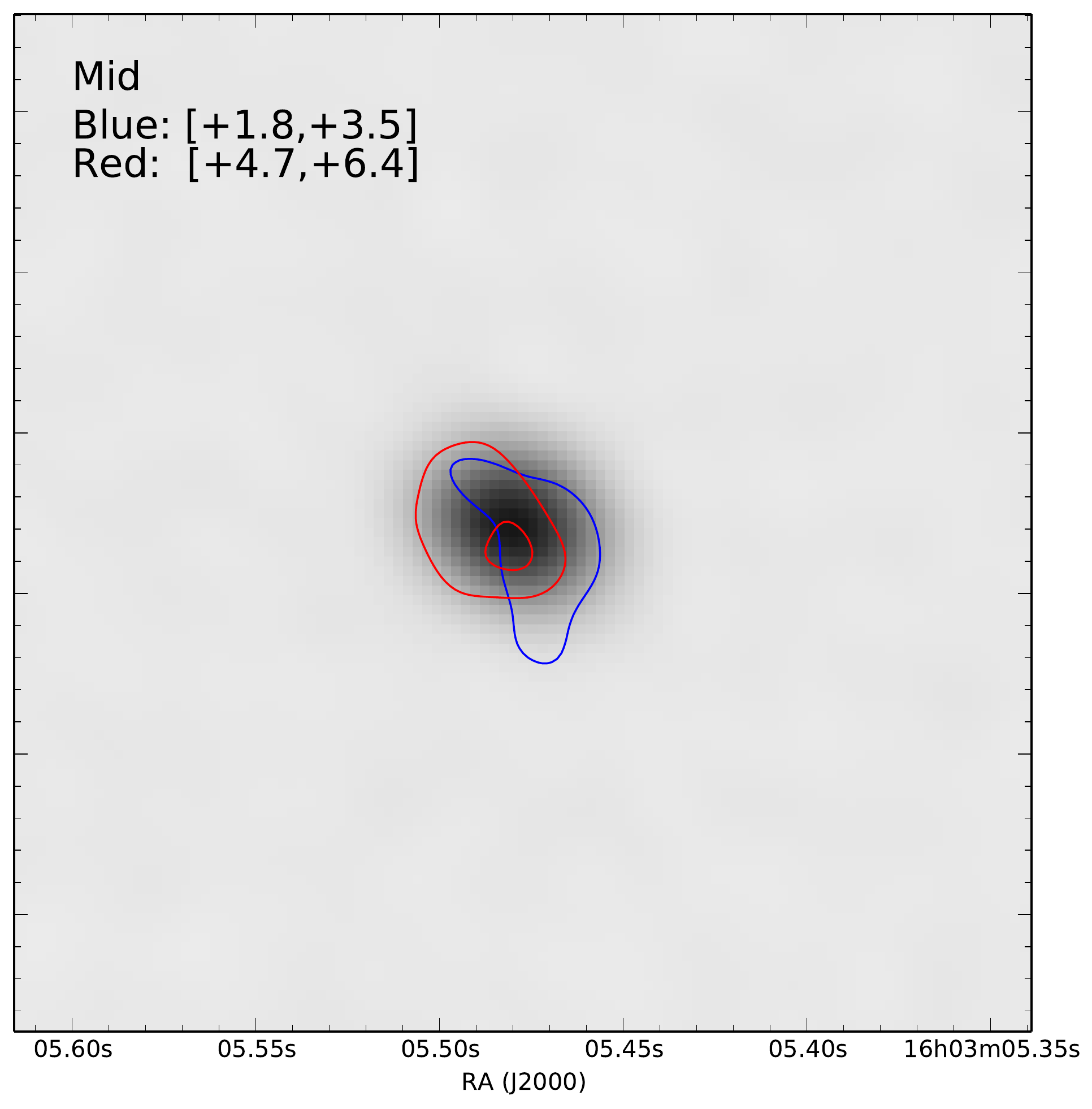}
\centering\includegraphics[angle=0,scale=0.25]{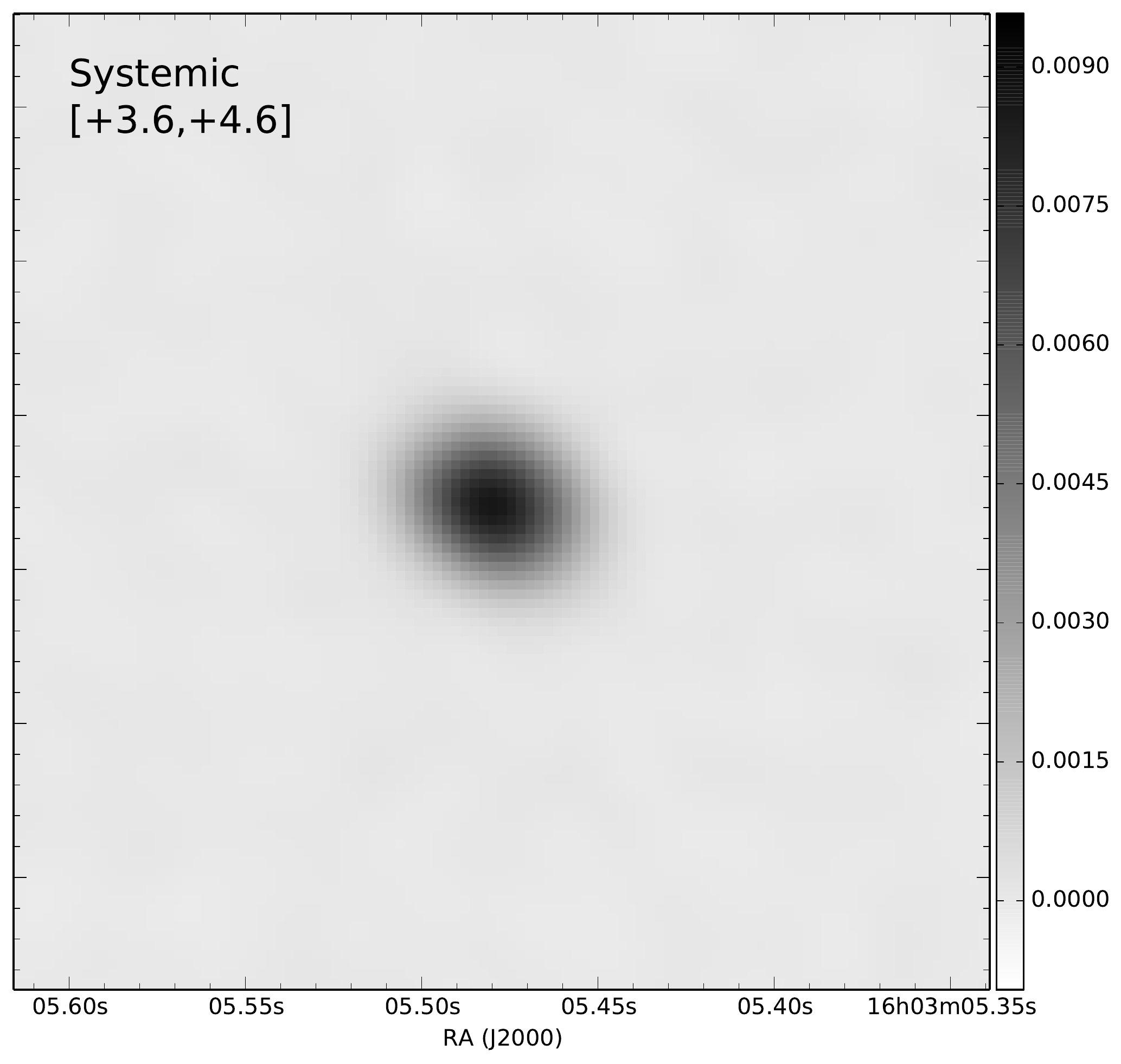}
\caption{Moment~0 maps in different velocity
  ranges. $^{13}$CO(2--1) and C$^{18}$O(2--1) are shown in the top
  and bottom rows, respectively. The high velocity channels ({\it{Left-panels}}) have been integrated
  between -1.6 and +1.7~km~s$^{-1}$ (blue) and between +6.5
  and +9.8~km~s$^{-1}$ (red). Intermediate velocity channels
  ({\it{Center-panels}}) have been integrated between +1.8 and
  +3.5~km~s$^{-1}$ (blue) and between +4.7 and
  +6.4~km~s$^{-1}$ (red). {\it Right-panels:} Integrated
  emission at systemic velocities (+3.6--+4.6~km~s$^{-1}$). Contour levels for $^{13}$CO and
  C$^{18}$O are same as in Figures~\ref{fig2_13co} and~\ref{fig2_c18o} respectively.}
\label{fig-mom-vels}
\end{figure}

Figures~\ref{fig2_13co} and ~\ref{fig2_c18o} show the channel maps for
$^{13}$CO(2--1) and C$^{18}$O(2--1), respectively. Figure~\ref{fig-mom-vels}
shows the corresponding moment~0 maps in different velocity ranges.

%
%
%
%
%
%

\end{document}